\documentclass[final,5p,times,twocolumn,authoryear]{elsarticle}
\usepackage{color}
\usepackage{amssymb}

\journal{}

\begin{document}

\begin{frontmatter}

\title{Scaling laws for the oblique impact cratering on an inclined granular surface}

\author[label1]{Shinta Takizawa}
\author[label1]{Hiroaki Katsuragi}

\address{Department of Earth and Environmental Sciences, Nagoya University, Nagoya 464-8601, Japan}

\begin{abstract}
Although a large number of astronomical craters are actually produced by the oblique impacts onto inclined surfaces, most of the laboratory experiments mimicking the impact cratering have been performed by the vertical impact onto a horizontal target surface. In previous studies on the effects of oblique impact and inclined terrain, only one of the impact angle $\varphi$ or target inclination angle $\theta$ has been varied in the experiments. Therefore, we perform impact-cratering experiments by systematically varying both $\varphi$ and $\theta$. A solid projectile of diameter $D_{\rm i}=6$~mm is impacted onto a sand surface with the range of impact velocity $v_{\rm i}=7$--$97$~m~s$^{-1}$. From the experimental result, we develop scaling laws for the crater dimensions on the basis of $\Pi$-group scaling. As a result, the crater dimensions such as cavity volume, diameter, aspect ratio, and depth-diameter ratio can be scaled by the factors $\sin \varphi$ and $\cos \theta$ as well as the usual impact parameters ($v_{\rm i}$, $D_{\rm i}$, density of projectile, and surface gravity). Finally, we consider the possible application of the obtained scaling laws to the estimate of impact conditions (e.g., impact speed and impact angle) in natural crater records. 
\end{abstract}


\end{frontmatter}

\section{Introduction}
Rocky astronomical bodies covered with regolith usually have a lot of impact craters on their surfaces. The majority of these craters has almost axisymmetric (circular) cavity structure. However, there are also asymmetric unusual craters. Possible origins of the unusual crater shapes are the oblique impact, topography, target heterogeneity, post-impact deformation (tectonic phenomena), or their combinations. Among them, target heterogeneity and tectonic effects are more or less geologic effects. In this study, we are interested in the instantaneous physical effects such as oblique impact onto inclined surface. For example, asymmetric ejecta deposition can be induced by the oblique impact~\citep{Melosh:1989,Melosh:2011}. In most of the natural impact events, the meteorite collides onto the target surface with an oblique angle. However, the population of asymmetric craters produced by oblique impacts on Mars is very limited~\citep{Herrick:2006}. On Earth, only one elliptical crater probably formed by the oblique impact has been found~\citep{Kenkmann:2009}. Actually, very shallow-angle impact is necessary to form asymmetric crater. Moreover, the critical angle to produce an elliptic crater depends on the size of crater~\citep{Collins:2011}. The effect of target inclination should also be considered to analyze asymmetric craters. Indeed, the non-circular craters have been observed on the inclined terrains of Moon~\citep{Plescia:2012,Neish:2014}, Mars~\citep{Aschauer:2017}, and asteroids~\citep{Elbeshausen:2012,Jaumann:2012,Krohn:2014}. On steeply sloped terrains, topography obviously influences the cratering process by modifying the transient crater shape due to the asymmetric landsliding driven by gravity.

To discuss the impact cratering on the regolith layer, various impact experiments on granular targets have been previously conducted. Particularly, in the granular physics field, crater morphology and penetration dynamics have been extensively studied by low-speed impact with impact velocity $v_{\rm i} \sim 10^0$~m~s$^{-1}$~\citep{Walsh:2003,Uehara:2003,Katsuragi:2007,Goldman:2008,Seguin:2009,Clark:2014}. Recently, even the ray-crater structure can also be reproduced by the laboratory granular impact experiments~\citep{Sabuwala:2018,PachecoVazquez:2019}. All of these experiments are the vertical impact onto a horizontal granular surface. In addition, there are some studies investigating the oblique impact onto a granular layer~\citep{Nishida:2010,Wang:2012}. While these studies have revealed the fundamental nature of granular impact phenomena, the applicability of the obtained physical laws to the astronomical impact cratering is not very clear. Particularly, the impact speed is much slower than the typical astronomical impact cratering.  

To directly mimic the astronomical oblique impact, \citet{Gault:1978} have conducted high-speed impact experiment ($v_{\rm i} \sim 10^3$~m~s$^{-1}$) by systematically varying the impact angle $\varphi$ using a solid projectile and quartz-sand target. Here, the impact angle $\varphi$ is defined by the angle from the flat surface, i.e., $\varphi=90^\circ$ corresponds to the impact perpendicular to the surface. 
In \citet{Gault:1978}, the ricochet of projectile was observed at shallow-angle impacts. The resultant craters possess the approximately circular shape when $\varphi$ is greater than $10^{\circ}$. However, the crater shape is elongated along the impact direction in the range of $\varphi\leq10^{\circ}$. 
The elongation degree depends on the type of projectile and impact velocity $v_{\rm i}$. 
The circularity of the crater shape significantly decreases only in the range of $\varphi\leq10^{\circ}$. However, the crater volume clearly depends on $\varphi$ even in relatively large $\varphi$ regime. Specifically, \citet{Gault:1978} have reported that the volume of crater cavity is proportional to $\sin\varphi$. Besides, the similar trend has been confirmed by three-dimensional numerical simulation~\citep{Elbeshausen:2009}. The scaling of crater volume which depends on the impact angle (in the range $30^{\circ} \leq \varphi\leq 90^{\circ}$) and friction coefficient of target materials has been numerically obtained by \citet{Elbeshausen:2009}.

The effect of inclined terrain has also been studied experimentally. In some previous studies, inclination of the target has been used to effectively mimic the oblique impact. However, the inclination itself must be an important factor governing the crater formation process. 
In general, astronomical bodies have various sloped terrains (e.g. large crater wall). Particularly, small bodies such as asteroids show large topographic slope variations. Therefore, the oblique impact experiments using only horizontal target are insufficient to fully understand the general impact cratering phenomena. Recently, solid-projectile-impact experiments using inclined granular target have been conducted~\citep{Hayashi:2017,Aschauer:2017,Takizawa:2019}. \citet{Hayashi:2017} performed the vertical free-fall-impact experiments in which the inclination angle of the dry granular surface $\theta$ and the impact kinetic energy $E$ are varied. They found that the resultant crater shapes can be divided into three phases. Obviously, a circular crater is formed by the vertical impact onto a horizontal sand target ($\theta=0^{\circ}$). The sharp rim structure is clearly left around the cavity (full-rim crater phase) in this type of circular crater. However, as $\theta$ increases, the final crater shape becomes shallower and elongated in the slope direction . In addition, the collapse of crater upper wall is induced at $\theta\geq 22^{\circ}$ (broken-rim crater phase). At the vicinity of repose angle of target sand ($\theta=34^{\circ}$), the crater cavity is almost buried by the large-scale avalanche towards the downslope direction (depression phase). \citet{Hayashi:2017} also found that the crater shape depends mainly on the inclination angle $\theta$ than the impact energy $E$. In other words, the scale of crater-wall collapse is principally determined by $\theta$. Similar trend has been confirmed by another previous study of the impact on an inclined granular surface~\citep{Aschauer:2017}. Although the range of impact speed is different between \citet{Hayashi:2017} ($\simeq 5$~m~s$^{-1}$) and \citet{Aschauer:2017} ($180$~m~s$^{-1}$), their results are very similar. \cite{Takizawa:2019} conducted the normal impact experiments on an inclined wet (cohesive) granular target and found that the catastrophic collapse of the slope can be induced by the impact when the target granular layer is cohesive enough and the inclination angle $\theta$ is close to the angle of repose. These previous studies suggest that the collapse of crater wall could significantly modify the crater shape when the target granular layer is inclined. Note that the inclination angle $\theta$ is defined relative to the gravitationally horizontal plane, i.e., horizontal surface corresponds to $\theta=0^{\circ}$. 

In the above-mentioned previous works, only one of the inclination angle $\theta$ or the impact angle $\varphi$ was varied. The crater formation process and the final crater shape depending on both $\theta$ and $\varphi$ have not yet been systematically clarified. In general astronomical impacts, meteorites obliquely collide onto sloped terrains. Nevertheless, $\theta$ and $\varphi$ have not been simultaneously varied in any laboratory experiment. Furthermore, the dimensionless scaling laws, which allow us to extrapolate the laboratory-experiment results to the astronomical impacts, have not been developed for the oblique impact onto an inclined surface. Dimensionless scaling called $\Pi$-group scaling~\citep{Buckingham:1914,Buckingham:1915} is a very powerful methodology to consider the scale-independent cratering dynamics. Indeed, the $\Pi$-group scaling has been applied to various impact-cratering analyses \citep{Schmidt:1980,Holsapple:1982,Holsapple:1987,Holsapple:1993,Holsapple:2007,Housen:2011}. However, the $\Pi$-group scaling has not been applied to the oblique impact onto an inclined surface.

\begin{figure}
\centering
\includegraphics[width=.5\linewidth]{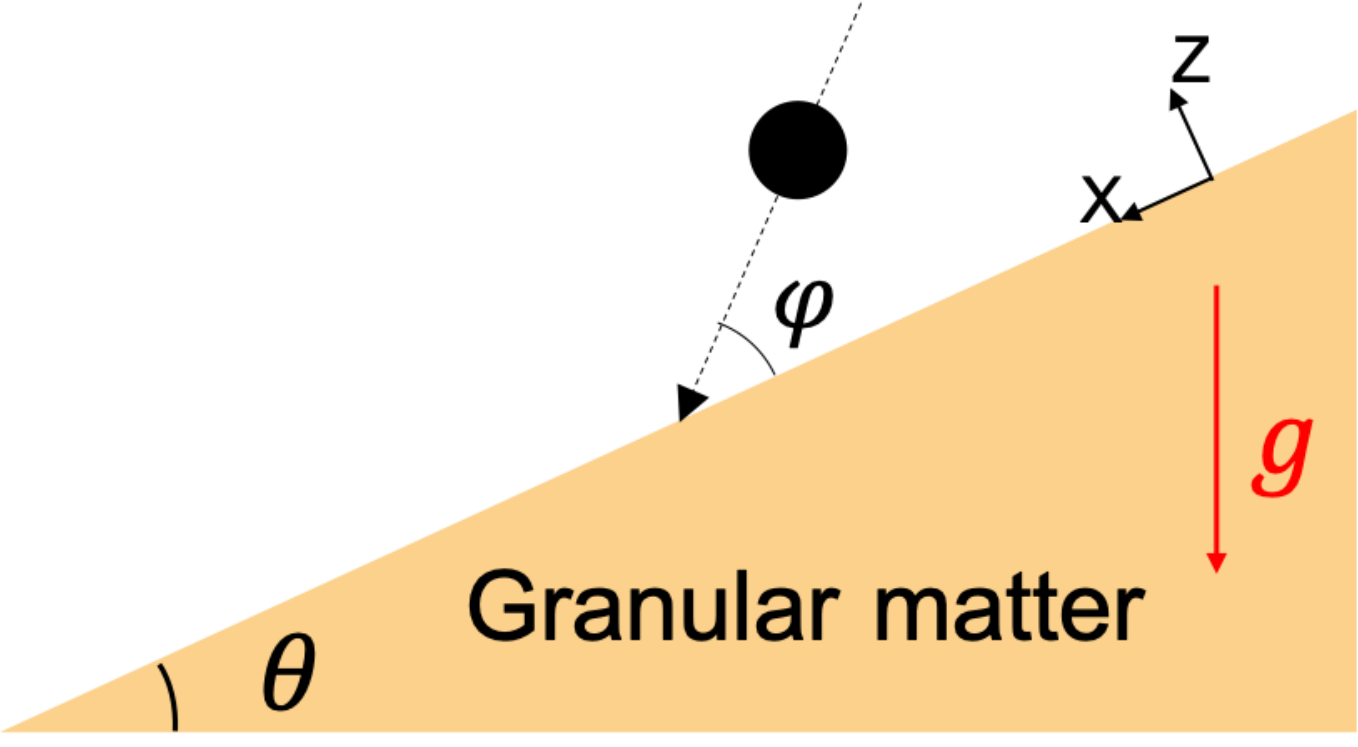}
\caption{Schematic diagram of the experiment and the definitions of the inclination angle $\theta$ and the impact angle $\varphi$. These angles are systematically and independently varied in this experiment.}
\label{fig:phi-theta-define}
\end{figure}

Therefore, in this study, we are going to develop the scaling laws for craters produced by the oblique impact onto an inclined granular surface.
To clarify the crater formation process and the scaling laws including the effects of $\theta$ and $\varphi$, we conduct experiments in which $\theta$ and $\varphi$ are systematically and independently varied~(Fig.~\ref{fig:phi-theta-define}). Then, we discuss the guideline for the possible application of the experimentally obtained scaling laws to astronomical impact-cratering analysis.

\section{Experiment}
\begin{figure}
\centering
\includegraphics[clip,width=1.0\linewidth]{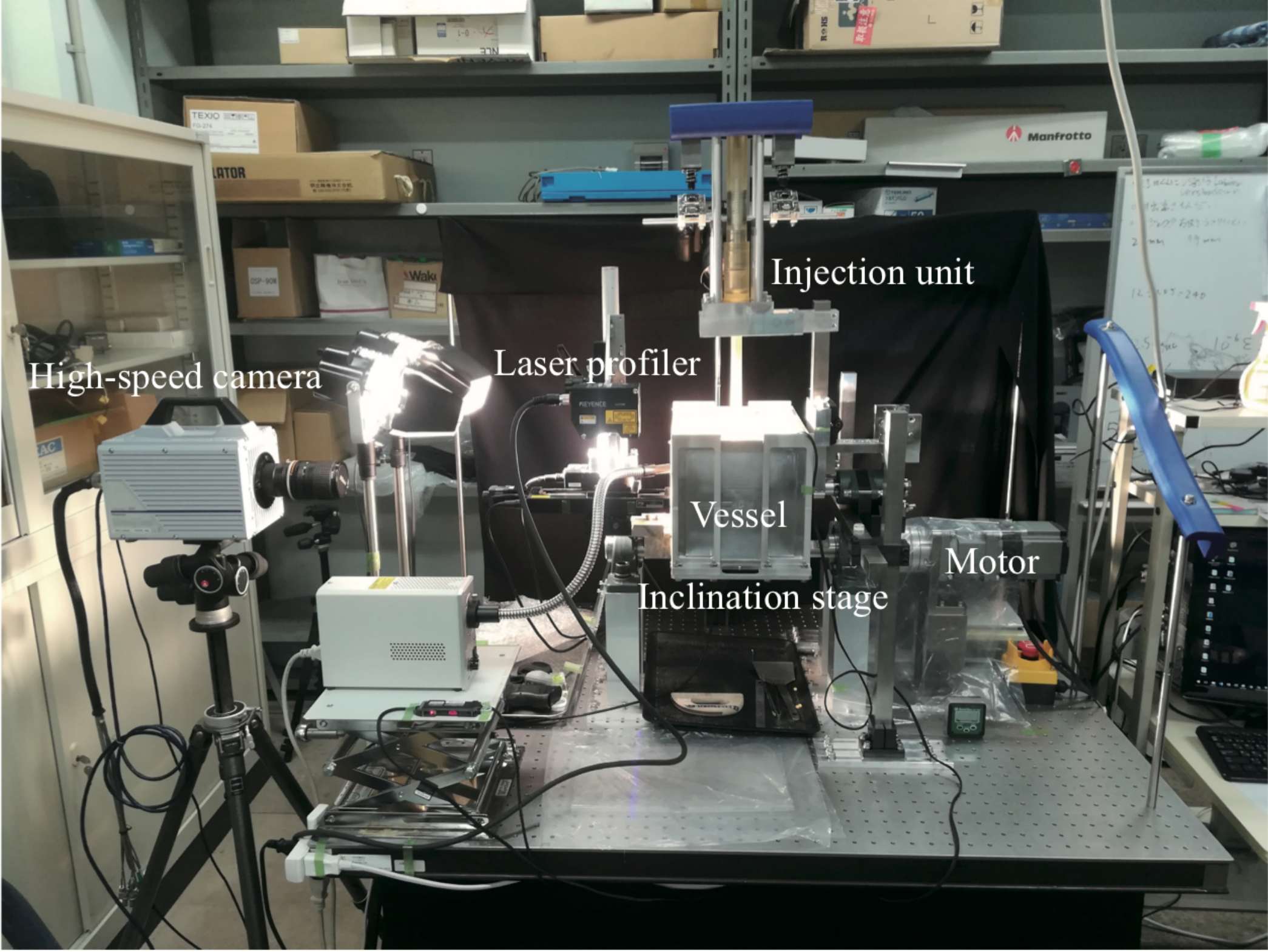}
\caption{
Photograph of the entire system of the experimental apparatus.
}
\label{fig:Experimental-apparutus}
\end{figure}

To perform the systematic impact experiments, we develop an experimental apparatus that can control both the inclination angle $\theta$ and the impact angle $\varphi$. The entire system of the developed experimental apparatus is shown in Fig.~\ref{fig:Experimental-apparutus}. The sand vessel is mounted on a tiltable inclination stage driven by a stepping motor (Orientalmotor, AZ98MCD-HS100). The resolution of rotation angle is $0.0036^{\circ}$ per pulse. Using this inclination stage, $\theta$ can be precisely controlled. The rotatable injection equipment (projectile gun) is also mounted on the inclination stage. The impact angle $\varphi$ is controlled by manually rotating the gun relative to the container vessel. Using a spring-based air-compression mechanism, a solid projectile (Tokyo Marui, BB-gun bullet) is injected towards the center of sand target. All experiments are performed under the atmospheric pressure condition. 

%% Figure:Experimental setup

\begin{figure}
\centering
\includegraphics[clip,width=1.0\linewidth]{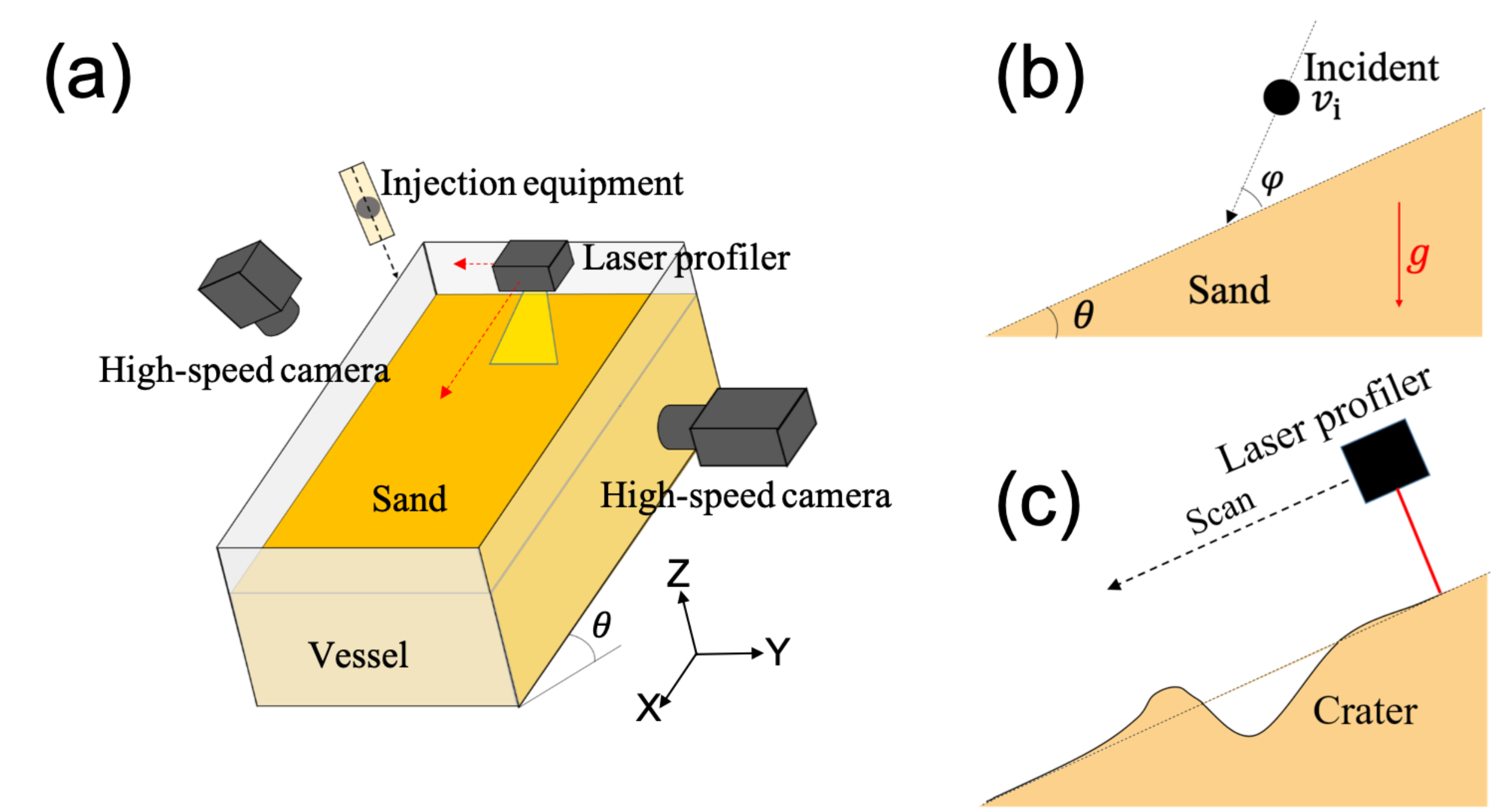}
\caption{
(a)~Schematic illustration of the experimental setup. Sand is poured into the container, and its surface is initially set to be parallel to the bottom of container by manual flattening. Then, the inclination angle $\theta$ is varied by rotating the stage. The impact angle $\varphi$ is controlled by rotating the injection unit. A two-dimensional (2D) laser profiler is attached to electronic stages  and scans the sand-target surface along the $XY$ plane. (b)~The projectile impacts onto the sand surface. Impact angle $\varphi$ and velocity $v_{\rm i}$ are measured by the high-speed camera. (c)~Three-dimensional (3D) profile of the sand-target surface is measured by the laser profiler before and after the projectile impact.
}
\label{fig:Schematic-of-setup}
\end{figure}

For target, Toyoura sand (TOYOURA KEISEKI KOGYO K.K.) is used. The diameter of sand grains ($0.1$--$0.3$~mm) is large enough to neglect the air-drag and humidity effects~\citep{Duran:2000,Andreotti:2013,Katsuragi:2016}. The true density and angle of repose of Toyoura sand are $2.63 \times10^3$~kg~m$^{-3}$ and $\theta_{\rm r}=34^{\circ}$, respectively. Thus, $\theta$ is varied in the range $0\leq \theta < \theta_r$. More detail physical properties of Toyoura sand can be found in \citep{Yamashita:2009}. We pour sand into a container (inner width:200~mm, length:300~mm, height:200~mm). The container width and length are sufficiently large to neglect the effect of the container wall on the impact dynamics~\citep{Seguin:2008,Nelson:2008}. To make frictional boundary, the identical sand grains are glued on the container walls. The thickness and packing fraction of the target sand layer is fixed to 100~mm and 0.55, respectively, in all experiments. First, the target layer is manually flattened at the horizontal position. Then, the target is tilted and the gun is rotated to control $\theta$ and $\varphi$. Since we use the common rotation axis ($Y$ axis), both angles $\theta$ and $\varphi$ are varied in the same two-dimensional space ($XZ$ plane). The variation of $\varphi$ in this experiment is in the range of $10^{\circ}\leq \varphi\leq170^{\circ}$. The $XYZ$ coordinate system is defined as shown in Fig.~\ref{fig:Schematic-of-setup}(a). The surface of target layer corresponds to $XY$ plane. 

%% Figure:Schematic of experimental setup

The injection gun shots a spherical projectile with diameter $D_{\rm i}=6$~mm and mass $m_{\rm i}=0.12$~g, 0.25~g, or 0.4~g. The range of impact speed is $7< v_{\rm i}\leq 97$~m~s$^{-1}$. The gun muzzle is kept at least 100~mm away from the target surface. The actual impact angle $\varphi$ and speed $v_{\rm i}$ are measured by using a high-speed camera (Photron, SA5) placed at the side of container, with a frame rate of 10,000 frames per second~(Fig.~\ref{fig:Schematic-of-setup}(b)). The crater formation process is taken by another high-speed camera (CASIO, EX-F1) placed in front of the sand surface, with a frame rate of 300 frames per second. The spatial resolutions of the side-view and the front-view images are 0.18~mm per pixel and 0.5~mm per pixel, respectively.

To measure the final crater shape formed by the impact, we use a linear two-dimensional (2D) laser profiler (KEYENCE, LJ-V7080) as shown in Fig.~\ref{fig:Schematic-of-setup}(c). This laser profiler, which measures the topography in width of $\sim 40$~mm, is attached to electronic stages (COMS, PM80B-200X, PM80B-100X, and PS60BB-360R) to slide the profiler in $XY$ plane. By combining line-profile data series, three-dimensional (3D) surface profile is synthesized. Since these electronic stages are mounted on the inclination stage, the crater profile along the surface of target (in $XY$ plane) can be obtained. The size of measurable $XY$ area is $191\times 65$~mm$^2$. The measurement resolution is 50~$\mu$m in horizontal ($XY$) direction and 0.5~$\mu$m in normal ($Z$) direction. These resolutions are sufficiently smaller than the mean diameter of sand grains 0.2~mm. 3D profiles of the sand surface are measured before and after the projectile impact, and the crater profile is computed by subtracting the before-impact profile from the after-impact profile. Using the obtained crater profiles, we measure the crater dimensions such as diameter, depth, and volume. 

The details of the developed experimental system will be reported elsewhere~\citep{Takizawa:2020}.

\section{Results}
\subsection{Crater shape dependence on inclination angle $\theta$ and impact angle $\varphi$}
\begin{figure}
\centering
\includegraphics[clip,width=1.0\linewidth]{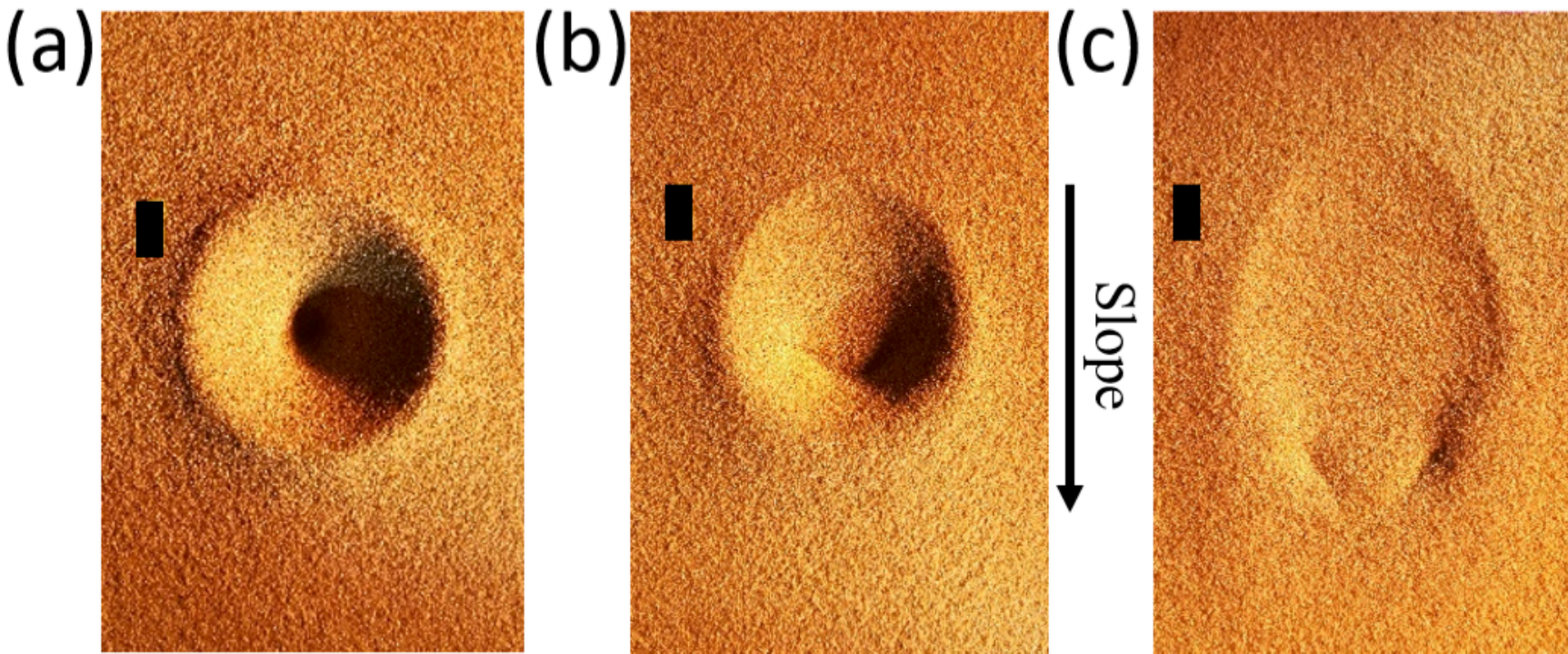}
\caption{
Final crater shapes produced by the impact conditions of (a)~$\theta=0^{\circ}$, $\varphi=90.2^{\circ}$, $v_{\rm i}=80.2$ m~s$^{-1}$, and $m_{\rm i}=0.12$~g, 
(b)~$\theta=20^{\circ}$, $\varphi=88.5^{\circ}$, $v_{\rm i}=77.5$~m~s$^{-1}$, and $m_{\rm i}=0.12$~g, and
(c)~$\theta=30^{\circ}$, $\varphi=90.5^{\circ}$, $v_{\rm i}=84.2$~m~s$^{-1}$, and $m_{\rm i}=0.12$~g.
Scale bars indicate 10~mm.
}
\label{fig:Experimental-craters}
\end{figure}

By the systematic impact experiments, morphology of resultant craters can qualitatively be classified. First, $\theta$ dependence of the crater shape is shown in Fig.~\ref{fig:Experimental-craters}. Figure~\ref{fig:Experimental-craters}(a) shows a symmetric circular crater formed by almost normal (and vertical) impact onto a horizontal sand surface ($\varphi=90^{\circ}$ and $\theta=0^{\circ}$). In Fig.~\ref{fig:Experimental-craters}(b,c), craters produced by the normal impact ($\varphi=90^{\circ}$) to the inclined sand surfaces are presented. As can be seen in Fig.~\ref{fig:Experimental-craters}(b,c), the asymmetry of the crater shape is enhanced as $\theta$ increases. This asymmetry mainly originates from the collapse of the wall of transient crater cavity. Particularly, the upper wall significantly collapses when the inclination angle $\theta$ is close to angle of repose. 

% Figure of the crater shapes

\begin{figure}
\centering
\includegraphics[clip,width=1.0\linewidth]{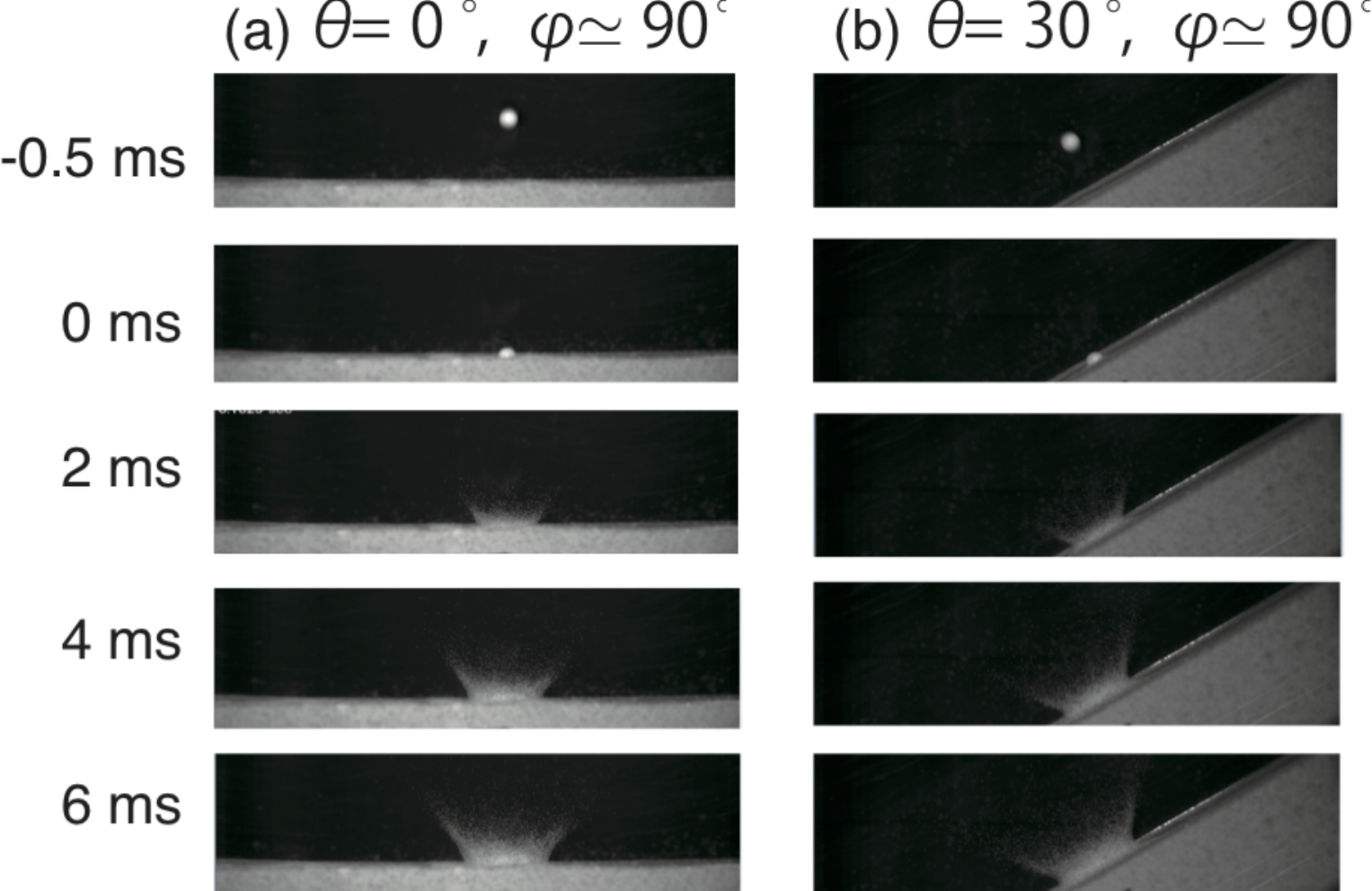}
\caption{
Sideview images of the impacts taken by a high-speed camera. Impact conditions are: (a)~$\theta=0^{\circ}$, $\varphi=90.2^{\circ}$, $v_{\rm i}=80.2$~m~s$^{-1}$, and $m_{\rm i}=0.12$~g, and 
(b)~$\theta=30^{\circ}$, $\varphi=88.5^{\circ}$, $v_{\rm i}=77.5$~m~s$^{-1}$, and $m_{\rm i}=0.12$~g. In both cases, the shape of ejecta curtain is almost symmetric around the normal axis to the surface.
}
\label{fig:Impact-mov}
\end{figure}

In Fig.~\ref{fig:Impact-mov}, high-speed images of the normal impact to (a)~horizontal and (b)~$30^\circ$-tilted sand surfaces are shown. One can confirm that the ejecta splashing right after the impact is almost isotropic in both cases. This means that the transient crater cavity should be almost axisymmetric around the normal axis at the impact point. Thus, to produce asymmetric craters shown in Fig.~\ref{fig:Experimental-craters}(b,c), the transient crater walls must collapse. Actually, the asymmetric collapse of the transient crater wall can be also observed in another high-speed video data as well. These observations are qualitatively consistent with recent previous experiments studying the cratering on inclined surfaces~\citep{Hayashi:2017,Aschauer:2017}.

% Figure of the crater shapes

\begin{figure}
\centering
\includegraphics[clip,width=1.0\linewidth]{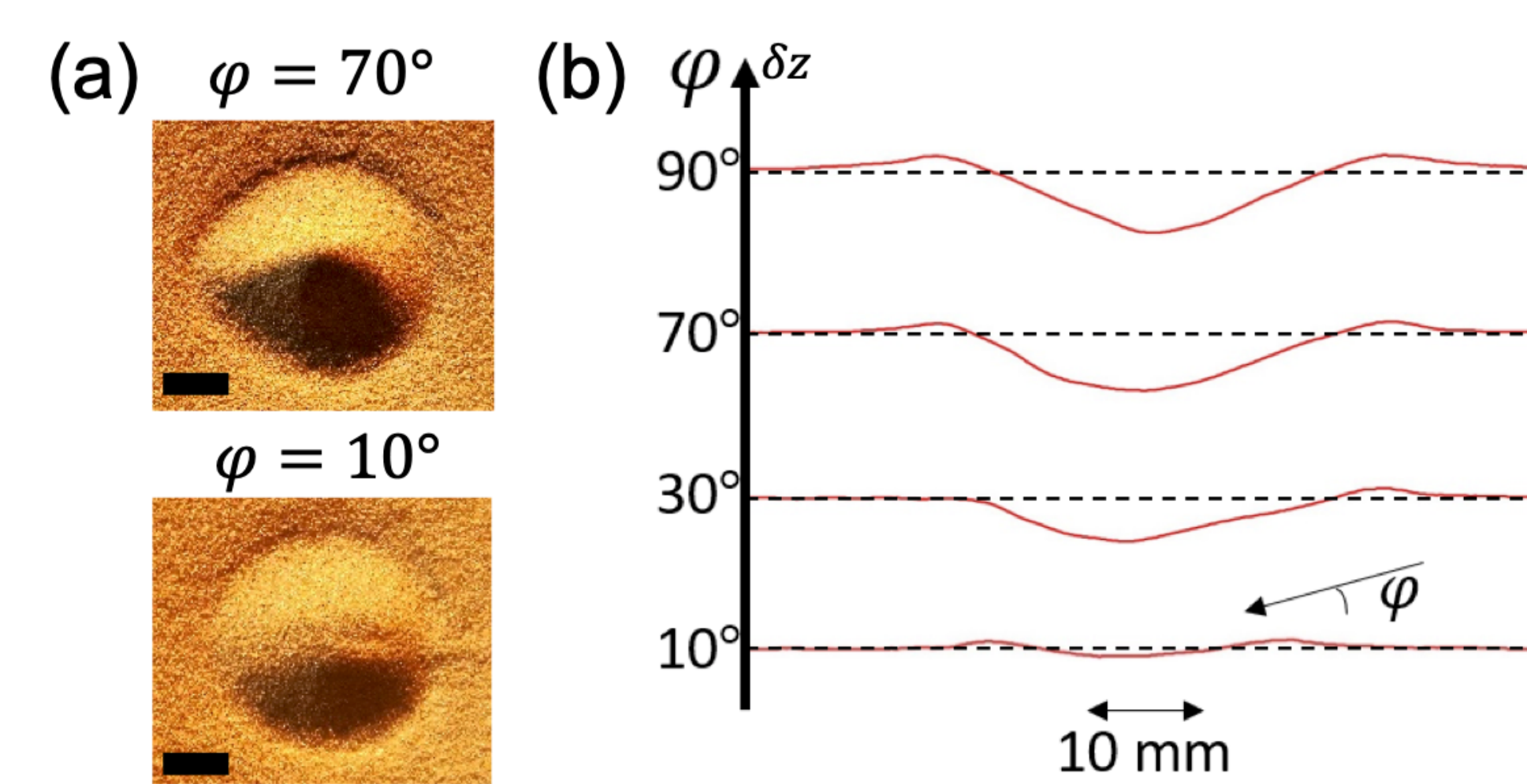}
\caption{
(a)~Photographs of craters formed by oblique impacts onto horizontal sand surfaces ($\varphi=70^{\circ}$, $10^{\circ}$ and $\theta=0^{\circ}$) are shown. Scale bars indicate 10~mm. Projectiles impact from the right and rebound to the left. 
(b)~The $\varphi$ dependence of cross-sectional crater profiles in $X$ direction. The crater shape becomes asymmetric and the cavity volume becomes small as $\varphi$ decreases. 
}
\label{fig:phi-dependence}
\end{figure}

In Fig.~\ref{fig:phi-dependence}, the effect of the impact angle $\varphi$ is presented. Figure~\ref{fig:phi-dependence}(a) shows crater shapes formed by oblique impacts onto horizontal sand surface ($\varphi\simeq70^{\circ}$ and $10^{\circ}$ with $\theta=0^{\circ}$). Due to the oblique impact, the crater shapes are slightly elongated. The crater wall seems to be partially removed by the projectile rebound when the rebound angle is shallower than the transient crater-wall angle. Note that the projectile comes from the right side and rebounds to the left side in Fig.~\ref{fig:phi-dependence}.  When $\varphi$ is in the range of $90\pm10^{\circ}$ (almost normal impact), rebound of projectile does not occur. However, the rebound of projectile can be observed in most of the oblique impacts. The rebound of projectile has also been confirmed in previous study of the oblique impact~\citep{Gault:1978}. Figure~\ref{fig:phi-dependence}(b) shows cross-sectional profiles in $X$ direction of the crater shape formed by oblique impact onto horizontal sand surfaces ($\varphi\simeq90$, $70$, $30$, and $10^{\circ}$ with $\theta=0^{\circ}$ and almost identical $v_{\rm i}$). The shallow, asymmetric, and small craters are produced by small $\varphi$ impacts.

%Figure of the oblique impact crater morphology

\subsection{Classification of crater wall collapse}
\begin{figure}
\centering
\includegraphics[clip,width=1\linewidth]{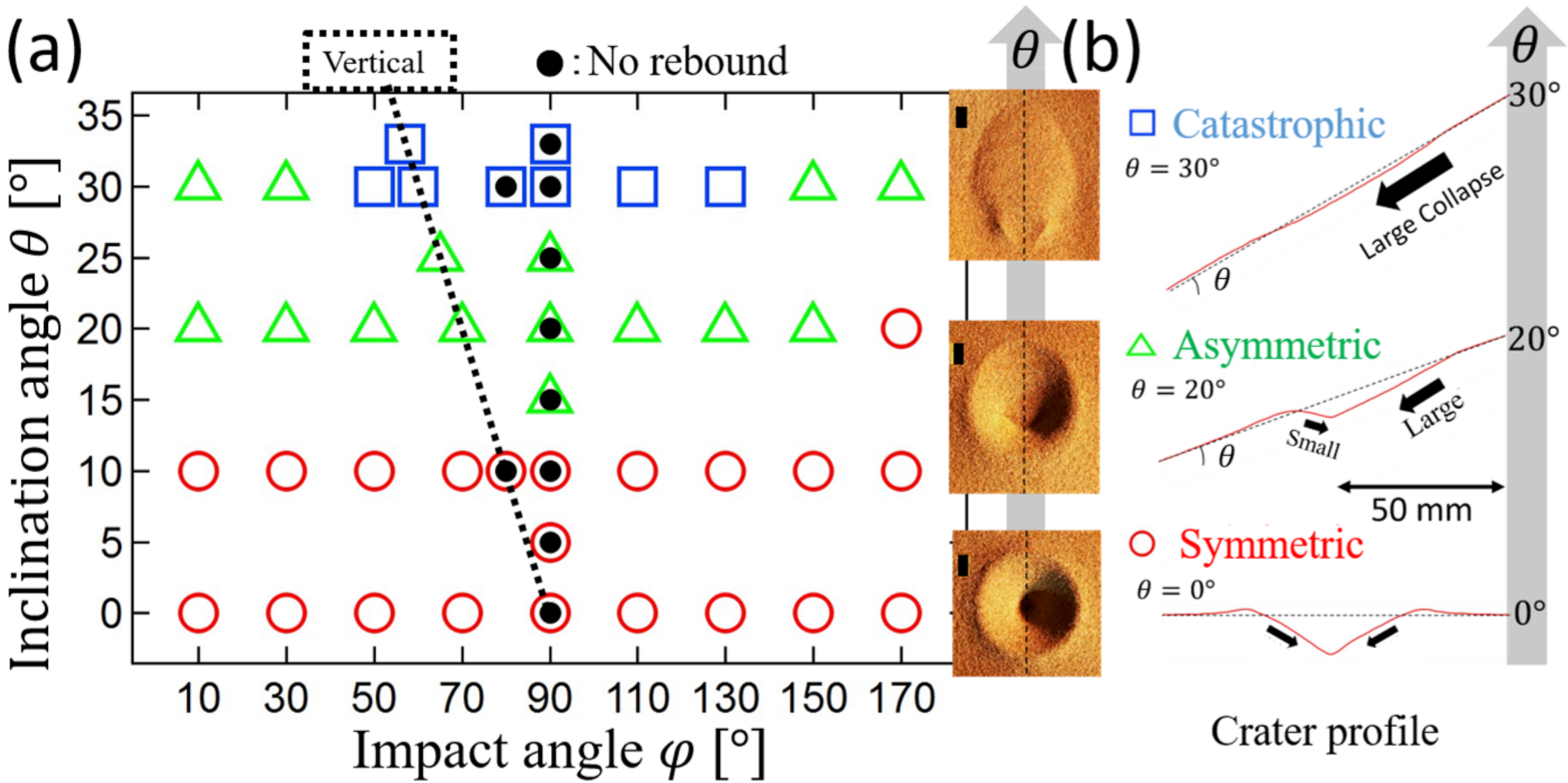}
\caption{
(a)~Classification diagram of the crater-wall collapse ($v_{\rm i}=10$--$100$~m~s$^{-1}$ and $m_{\rm i}=0.12$~g). Three modes (symmetric, asymmetric, and catastrophic collapse) are mainly determined by $\theta$ and almost independent of $\varphi$. A broken line indicates the vertical impact. The black filled circles in the symbols indicate that the projectile does not rebound. The projectile rebounds in all the hollow-symbol cases. (b)~Examples of crater shapes and corresponding cross sections in $X$ direction for each mode are shown. The data with $\varphi=90^{\circ}$ and $\theta=0^{\circ}$, $20^{\circ}$, and $30^{\circ}$ are shown. Scale bars in the photos indicate $10$~mm.
}
\label{fig:Phase-diagram}
\end{figure}

By systematically varying $\theta$ and $\varphi$, we observe three representative modes of crater wall collapse (symmetric, asymmetric, and catastrophic collapses). The symmetric-collapse mode is defined by the case in which the transient crater cavity formed by the excavation collapses isotropically (symmetrically). This modification is very small so that the initial transient cavity is almost preserved. On the other hand, asymmetric-collapse mode is characterized by the asymmetric collapse of upper wall of the transient crater cavity which results in the asymmetric final crater shape. In addition, the catastrophic collapse mode is defined by the large-scale collapse in which the flow initiated at the upper crater wall reaches the lower crater rim. Figure~\ref{fig:Phase-diagram} shows the collapse-mode diagram based on the above-mentioned classification of the transient-crater-wall collapse. The diagram is independent of $v_{\rm i}$ in the experimented range $\sim 10$--$100$~m~s$^{-1}$. As confirmed in Fig.~\ref{fig:Phase-diagram}, the scale of the collapse increases as $\theta$ increases. The obtained classification diagram is qualitatively consistent with the previous study~\citep{Hayashi:2017}. Moreover, one can also confirm that the collapse scale is hardly influenced by the impact angle $\varphi$. However, the rebound condition seems to depend on $\varphi$. Black filled circles in Fig.~\ref{fig:Phase-diagram}(a) indicate no-rebound cases. As shown in Fig.~\ref{fig:Phase-diagram}(a), most of the oblique impacts result in the projectile rebound. Although the physical criterion for the rebound conditions is not clearly understood, the detail analysis of the rebound speed and angle will be presented elsewhere ~\citep{Takizawa:2020}. In Fig.~\ref{fig:Phase-diagram}(b), the typical final crater shapes and the corresponding crater profiles (in $X$ direction) are shown. As $\theta$ increases, the asymmetry is enhanced, and finally the crater cavity is almost buried by the significant upper crater-wall collapse at $\theta=30^{\circ}$.

%Figure phase diagram

\begin{figure}
\centering
\includegraphics[clip,width=1.0\linewidth]{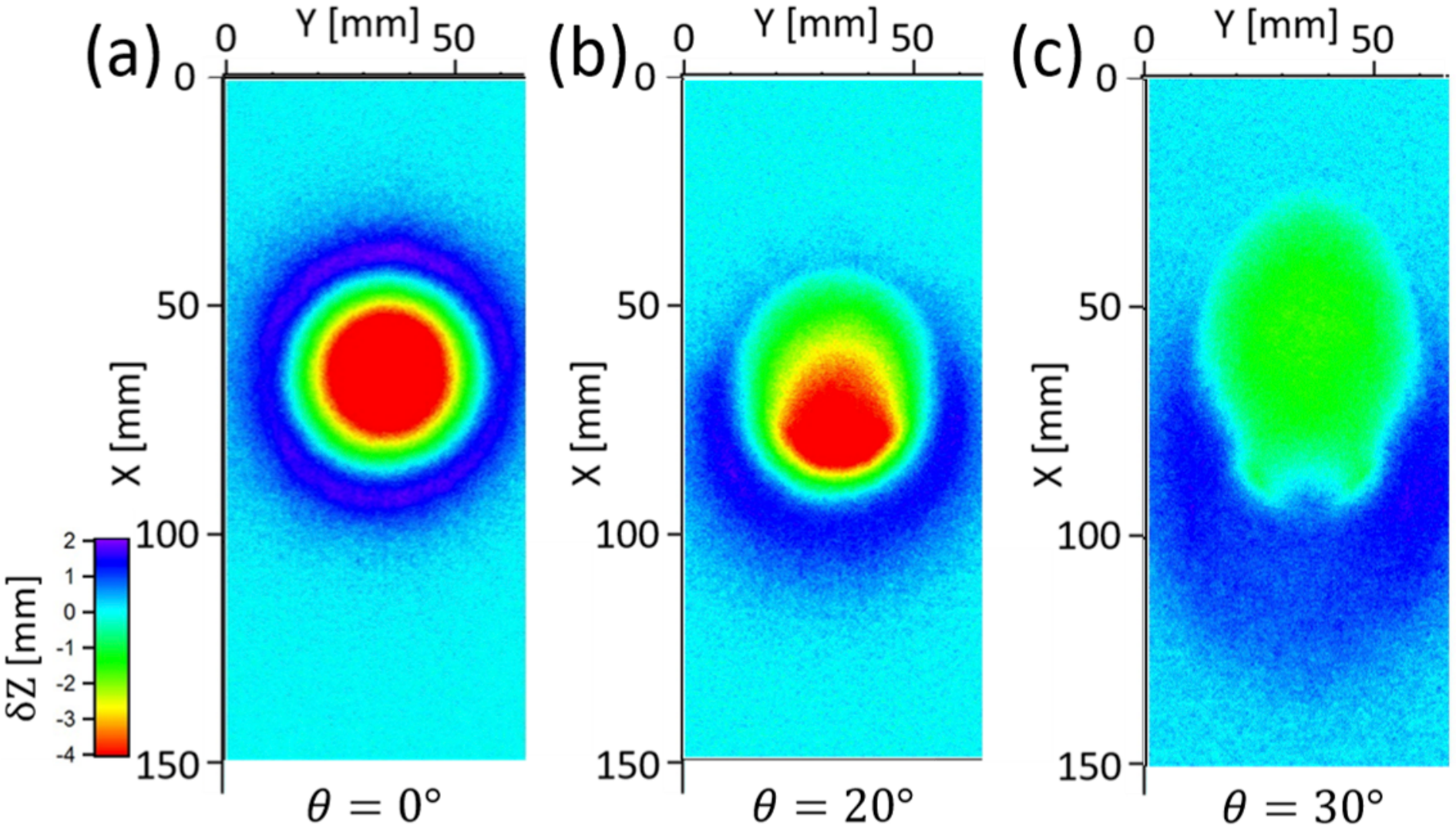}
\caption{
Three-dimensional profiles of the final craters produced by the impact conditions of (a)~$\theta=0^{\circ}$, $\varphi=90.2^{\circ}$, $v_{\rm i}=80.2$ m~s$^{-1}$, and $m_{\rm i}=0.12$ g, 
(b)~$\theta=20^{\circ}$, $\varphi=88.5^{\circ}$, $v_{\rm i}=77.5$~m~s$^{-1}$, and $m_{\rm i}=0.12$ g, and 
(c)~$\theta=30^{\circ}$, $\varphi=90.5^{\circ}$, $v_{\rm i}=84.2$~m~s$^{-1}$, and $m_{\rm i}=0.12$ g. 
The origin of the plots corresponds to the top-left corner of the measured region, and $\varphi$ is defined as shown in Fig.~\ref{fig:phi-theta-define}.
}
\label{fig:Crater-profile}
\end{figure}

Examples of 3D crater profile measured by the laser profiler are shown in Fig.~\ref{fig:Crater-profile}. The crater depth $\delta Z$ indicates the height difference between before and after the impact. In Fig.~\ref{fig:Crater-profile}, $\theta$ is varied while $\varphi=90^{\circ}$ is fixed (i.e., normal impacts). At $\theta=0^{\circ}$, the wall of the transient crater slightly collapses in a isotropic way so that the isotropic rim clearly exists around the crater cavity (Fig.~\ref{fig:Crater-profile}(a)). In contrast, for the cases of $\theta\geq 20^{\circ}$ (Fig.~\ref{fig:Crater-profile}(b,c)), the upper rim structure is flown out due to the collapse of the upper wall. In addition, the deepest point in the cavity migrates downward compared to $\theta=0^{\circ}$, causing the asymmetric profiles as also shown in Fig.~\ref{fig:Phase-diagram}(b). That is, both the rim and cavity structures are modified by the effect of inclination~$\theta$. At $\theta=30^{\circ}$, the crater cavity has an unusual (almost flat) shape because the large-scale collapse of the upper crater wall reaches the lower crater rim (Fig.~\ref{fig:Crater-profile}(c)). These trends are qualitatively consistent with previous studies~\citep{Hayashi:2017,Aschauer:2017}. The effect of impact angle $\varphi$ on the crater asymmetry is actually limited compared to the effect of $\theta$. As shown in Fig.~\ref{fig:phi-dependence}, the small $\varphi$ results in the small crater cavity rather than the enhancement of asymmetry. 

\subsection{Definition of crater dimensions}
\begin{figure}
\centering
\includegraphics[clip,width=1.0\linewidth]{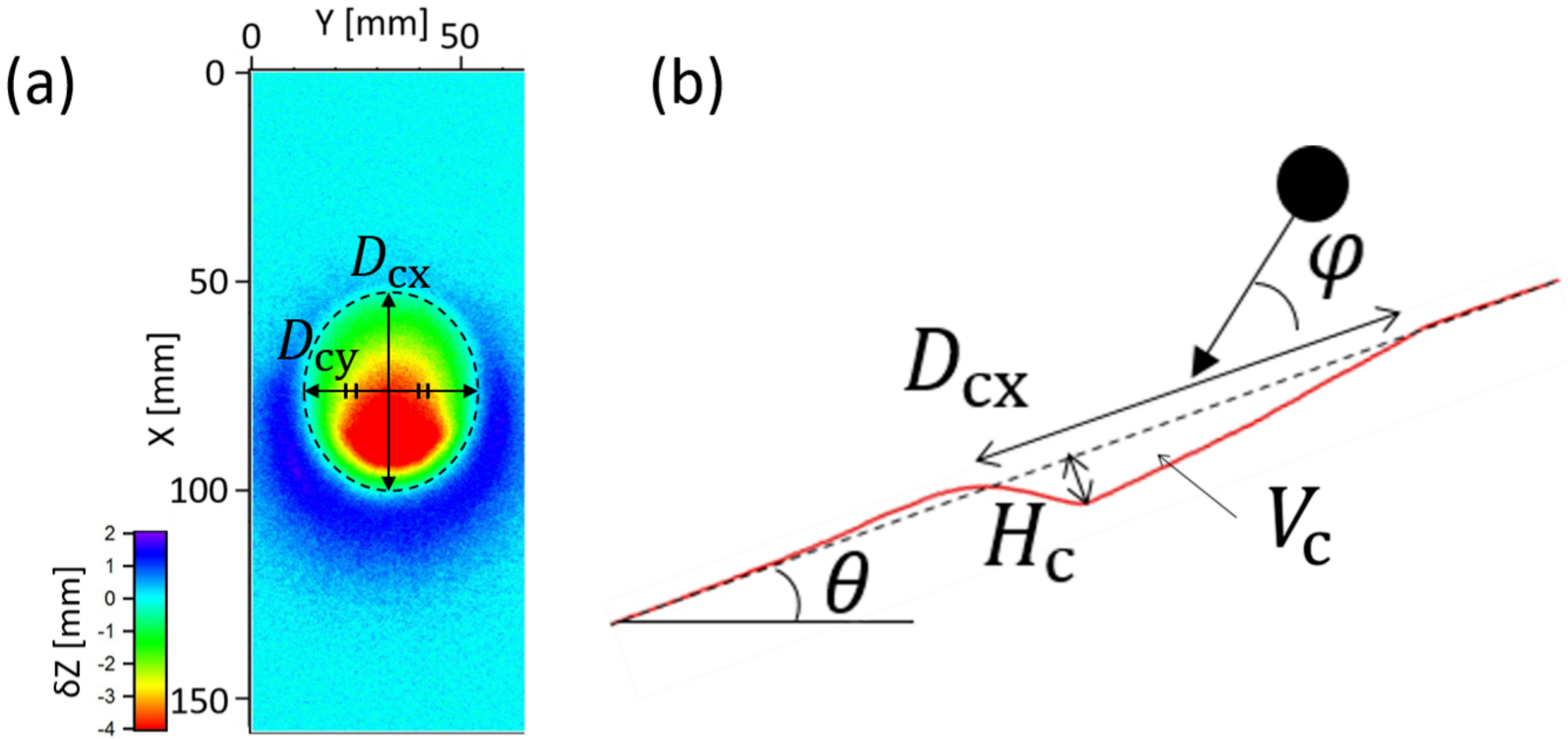}
\caption{
(a)~An example of 3D crater profile with definitions of $D_{\rm cx}$ and $D_{\rm cy}$. Almost elliptical broken curve indicates ${\delta}Z=0$ defining the crater cavity perimeter. The crater width $D_{\rm cy}$ and the crater length $D_{\rm cx}$ correspond to the diameters in $Y$ and $X$ directions, respectively. (b)~The cross-sectional profile taken along a line of $D_{\rm cx}$. A broken line indicates the level of original surface ($\delta Z=0$). The crater depth $H_{\rm c}$ is defined by the largest negative displacement ($\max{|\delta Z|}$). 
}
\label{fig:Definitions-craterdimensions}
\end{figure}

For quantitative analyses of the final crater shape, the following crater dimensions are defined and measured: length $D_{\rm cx}$, width $D_{\rm cy}$, depth $H_{\rm c}$, and volume $V_{\rm c}$. Fig.~\ref{fig:Definitions-craterdimensions} shows an example of (a)~3D crater profile and (b)~corresponding cross-sectional profile. A broken circular curve in Fig.~\ref{fig:Definitions-craterdimensions}(a) indicates a contour of $\delta Z=0$ (around the crater floor) defining the outline of crater cavity. The crater width $D_{\rm cy}$ is the maximum width of the contour ($\delta Z=0$) in $Y$ direction (perpendicular to the inclination direction). The crater length $D_{\rm cx}$ is the linear dimension of the contour ($\delta Z=0$) in $X$ direction at the center of $D_{\rm cy}$. Usually, $D_{\rm cx}$ and $D_{\rm cy}$ correspond to the major and minor axes of the crater shape. The crater depth $H_{\rm c}$ is defined by the largest negative displacement ($\max{|\delta Z|}$) in the profile (Fig.\ref{fig:Definitions-craterdimensions}(b)). The crater volume $V_{\rm c}$ is the volume of the crater cavity ($\delta Z\leq0$). In the following subsections, the measured data will be presented and analyzed. All the measured data and corresponding impact conditions are listed in the supplementary data file (all-data.csv).

\subsection{Energy dependence of the crater dimensions}
\begin{figure}
\centering
\includegraphics[clip,width=1.0\linewidth]{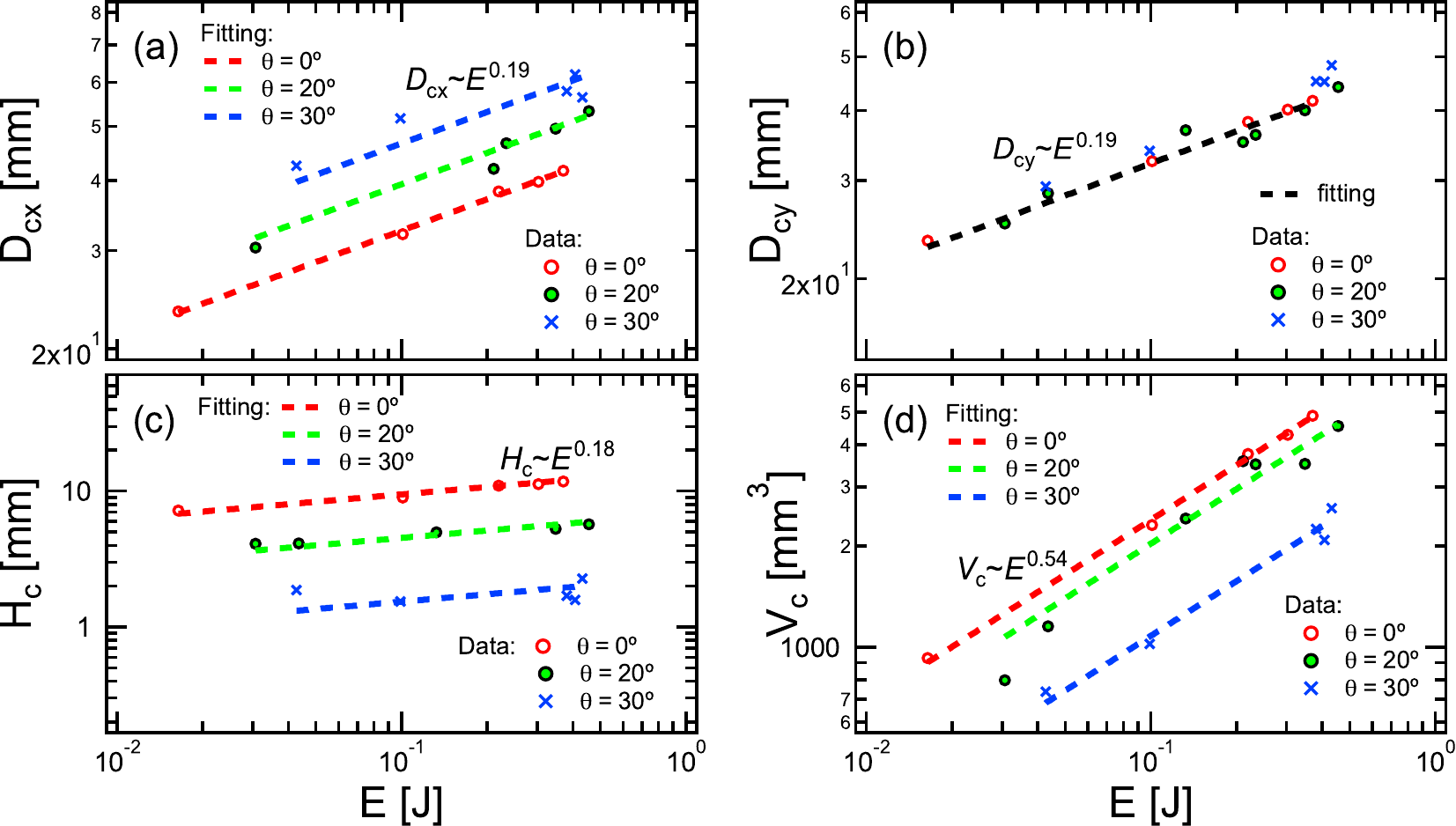}
\caption{
Impact energy $E$ dependence of the (a)~crater length $D_{\rm cx}$, (b)~crater width $D_{\rm cy}$, (c)~crater depth $H_{\rm c}$, and (d)~crater volume $V_{\rm c}$ under the conditions of $\varphi=90^{\circ}$ and $m_{\rm i}=0.12$~g. Open circular (red), filled circular (green), and cross (blue) symbols indicate the inclination angles $\theta$= 0, 20, and $30^{\circ}$, respectively. Broken lines are the scaling relations.
}
\label{fig:Energy-scalings}
\end{figure}

To understand the crater formation process, we investigate the relation among impact kinetic energy and the crater dimensions defined above. Here, the impact kinetic energy $E$ is simply defined as
\begin{equation}
 E=\frac{1}{2}m_{\rm i}{v_{\rm i}}^2.
 \label{eq:E}
\end{equation}
 Figure~\ref{fig:Energy-scalings} shows $E$ dependence of the crater dimensions $D_{\rm cx}$, $D_{\rm cy}$, $H_{\rm c}$, and $V_{\rm c}$ for various $\theta$ (with fixed $\varphi=90^{\circ}$). As seen in Fig.~\ref{fig:Energy-scalings}, all of these crater dimensions show power-law relations with $E$. The broken lines in Fig.~\ref{fig:Energy-scalings} indicate the power-law fittings. The obtained scaling exponents for $D_{\rm cx}$, $D_{\rm cy}$, and $H_{\rm c}$ almost coincide with each other for all inclination-angle ($\theta$) cases. The value of scaling exponent is approximately obtained as $0.19$. The scaling of crater diameter has been studied well by the low-speed granular impact using solid and liquid-drop projectiles. The obtained scaling exponent is basically close to $1/4=0.25$~\citep{Walsh:2003,Uehara:2003,Katsuragi:2010}. However, the scaling exponent $1/6=0.17$ has also been reported in a droplet impact onto a granular layer~\citep{Zhao:2015}. The values obtained in this study ($0.18$--$0.19$; Fig.~\ref{fig:Energy-scalings}(a-c)) are close to $1/6$. The scaling exponent for crater volume $V_{\rm c}$ is about ${0.54}$ (Fig.~\ref{fig:Energy-scalings}(d)). This value can roughly be derived as $V_{\rm c} \sim D_{\rm cx}D_{\rm cy}H_{\rm c} \sim E^{0.19+0.18+0.19} \sim E^{0.56}$. This exponent value actually relates to the the scaling obtained by $\Pi$-group scaling ($\pi_2^{-0.52}$ in Eq.~(\ref{eq:piv-simple})) and is almost consistent with previously obtained one (see Sec.~\ref{sec:scaling-crater-dimensions}). Although the scaling exponents seem to be universal, the specific values (scaling coefficients) of $D_{\rm cx}$, $H_{\rm c}$, and $V_{\rm c}$ depend on $\theta$. In contrast, $D_{\rm cy}$ is independent of $\theta$. All of the $D_{\rm cy}$ data of various $\theta$ cases are scaled by the single power-law relation (Fig.~\ref{fig:Energy-scalings}(b)). These characteristics ($\theta$-independent $D_{\rm cy}$ and $\theta$-dependent $D_{\rm cx}$ and $H_{\rm c}$) are found for the first time in this experiment. Moreover, we find that all the exponents are insensitive to $\theta$. The value of $\theta$ only affects the coefficients in the scaling relations. 

\section{Analysis and discussion}
\subsection{Crater modification process}
\begin{figure}
\centering
\includegraphics[clip,width=1.0\linewidth]{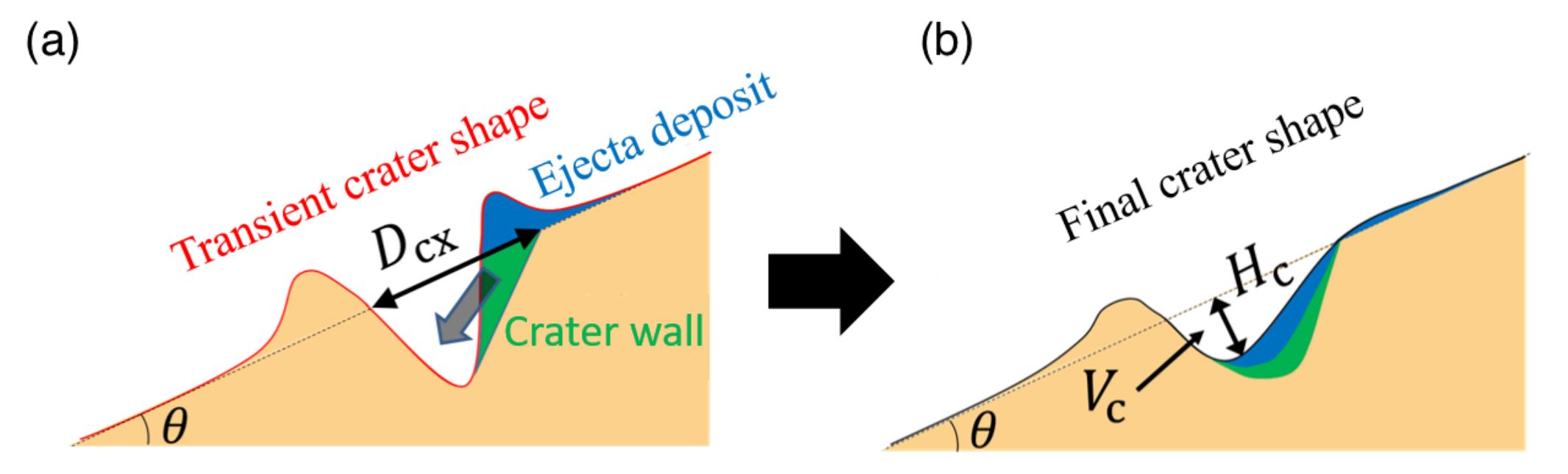}
\caption{
Schematic illustration of the transient-crater collapse (modification) process due to the effect of inclination angle $\theta$. (a)~Transient crater shape before the upper-wall collapse is drawn by red curve. The crater length $D_{\rm cx}$ is an increasing function of $\theta$ because the scale of collapse of crater wall and ejecta deposit (green and blue regions in (a)) increases with $\theta$. (b)~The final crater shape is formed by the reaccumulation of the collapsed crater wall and ejecta deposit falling back into the crater cavity. Due to this process, the crater depth $H_{\rm c}$ and the crater volume $V_{\rm c}$ are decreasing functions of $\theta$.
}
\label{fig:collapse-process}
\end{figure}

From these observations, the collapse process of transient crater wall can be qualitatively understood. Figure~\ref{fig:collapse-process}(a) shows a schematic image of the transient crater shape (cross section). The collapse of upper crater wall and deposited ejecta (green and blue regions, respectively, in Fig.~\ref{fig:collapse-process}(a)) is triggered when the slope of upper crater wall is steep enough to be unstable. The scale of collapse is mainly determined by the inclination angle $\theta$. As a result, $D_{\rm cx}$ becomes an increasing function of $\theta$. The final crater shape is formed by the reaccumulation of the collapsed crater wall and ejecta deposits falling back into the transient crater cavity (Fig.~\ref{fig:collapse-process}(b)). Thus, the above-mentioned collapse mechanism is also consistent with the negative correlation between $H_{\rm c}$ and $\theta$. Furthermore, due to the ejecta volume falling back to the crater cavity, the crater volume $V_{\rm c}$ becomes a decreasing function of $\theta$ (Fig.~\ref{fig:collapse-process}(b)). However, the crater width $D_{\rm cy}$ is independent of $\theta$ because $D_{\rm cy}$ is perpendicular to the collapse direction. While these interpretations are consistent with the measured results, it is difficult to directly analyze the time-resolved dynamics of the collapse of transient crater cavities. The temporal and spatial resolutions of the front-view movies are quite insufficient to analyze the dynamics. Detail characterization of the dynamics in crater-collapse process by more precise measurement method is an important future work. Instead, we focus on the analysis of final crater dimensions. Quantitative scaling analysis is reported in the next subsection.

\subsection{Scaling of the crater dimensions}
\label{sec:scaling-crater-dimensions}
For further quantitative understanding of the cratering process, scaling with dimensional analysis is employed here. The appropriate dimensional analysis is necessary to discuss the applicability of the laboratory-experiment results to astronomical-scale phenomena. Here we use the standard $\Pi$-group scaling to analyze the current experimental results. 

Because the final crater volume $V_{\rm c}$ is determined by the combination of all effects, we first develop the scaling law for $V_{\rm c}$. Based on the $\Pi$-group scaling, we start with two dimensionless numbers: 
\begin{equation}
\pi_{\rm v}=\frac{\rho_{\rm t}V_{\rm c}}{m_{\rm i}} \;\mbox{ and }\; \pi_{\rm 2}=\frac{gD_{\rm i}}{v_{\rm i}^2},
 \label{eq:dimensionless numbers}
\end{equation} 
 where $\rho_{\rm t}$, $V_{\rm c}$, $m_{\rm i}$, $g$, $D_{\rm i}$, and $v_{\rm i}$ are the density of target, crater volume, projectile mass, gravity of the impacted body, projectile diameter, and impact speed, respectively. Since the target consists of cohesionless sand, effective strength of the target material is negligibly small. In such situation, surface gravity dominates the cratering dynamics. In this gravity-dominant regime, $\pi_{\rm 2}$ is a relevant dimensionless parameter. $\pi_{\rm v}$ of the craters produced by the vertical impact onto a horizontal sand surface is well scaled by $\pi_{\rm 2}$~\citep{Holsapple:1987}. Thus, we first examine the relation between crater efficiency (normalized crater volume) $\pi_{\rm v}$ and the gravity-scaled size $\pi_{\rm 2}$. 

\begin{figure}
\centering
\includegraphics[clip,width=1.0\linewidth]{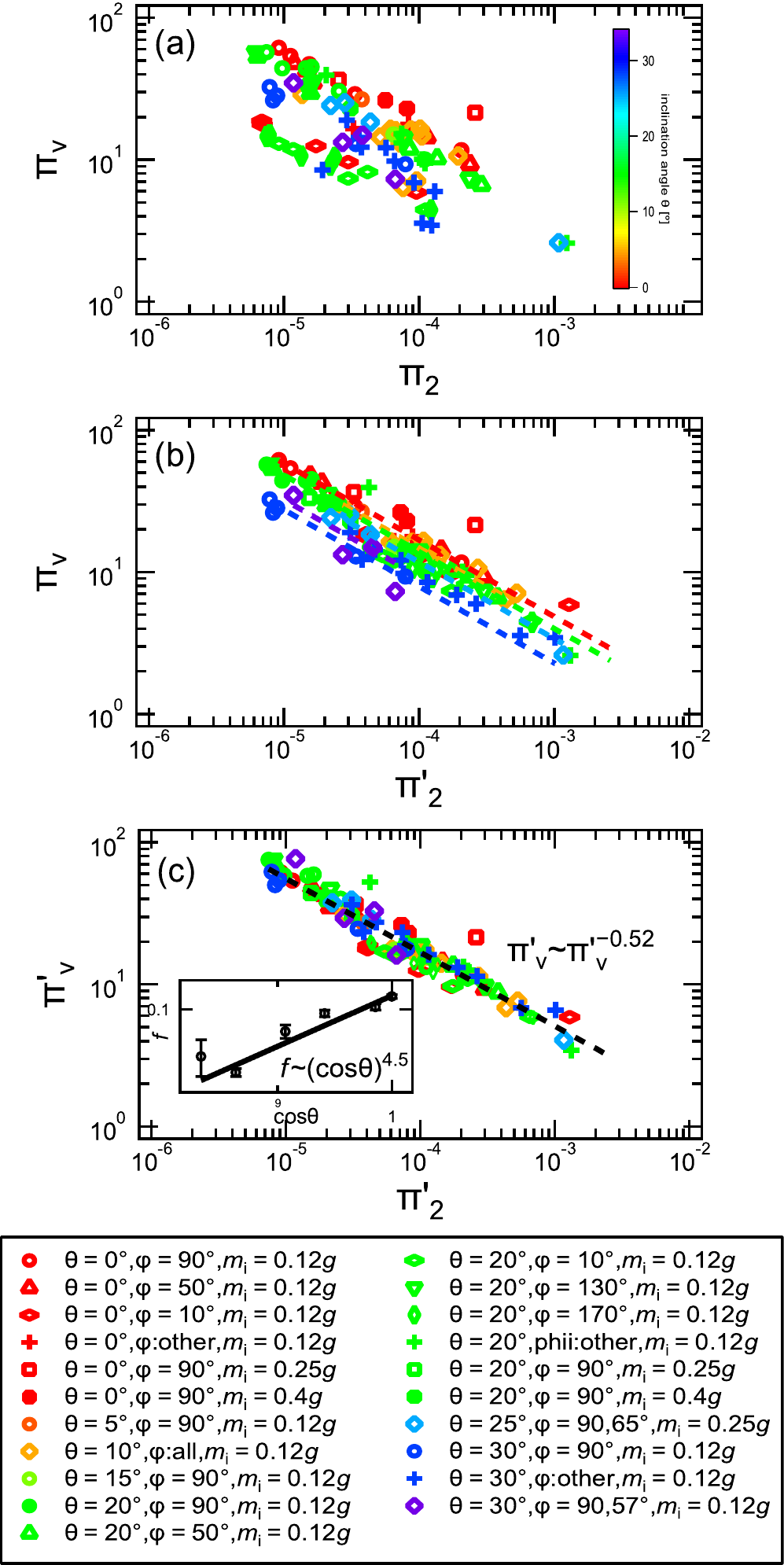}
\caption{
Scaling of the normalized crater volume. 
(a)~The gravity-scaled size $\pi_{\rm 2}$ dependence of the cratering efficiency $\pi_{\rm v}$.
(b)~$\pi'_{\rm 2}$~(Eq.~(\ref{eq:pi2'})) dependence of $\pi_{\rm v}$ taking into account the impact angle $\varphi$.
 (c)~$\pi'_{\rm 2}$ dependence of $\pi'_{\rm v}$~(Eq.~(\ref{eq:piv'_difinition})) taking into account the inclination angle $\theta$. The inset of (c) shows $\cos\theta$ dependence of $f(\theta)$.
}
\label{fig:cratering-efficiency}
\end{figure}

 Figure~\ref{fig:cratering-efficiency}(a) shows $\pi_{\rm 2}$ dependence of $\pi_{\rm v}$. All experimental data with various $\theta$, $\varphi$, $v_{\rm i}$, and $m_{\rm i}$ are plotted in Fig.~\ref{fig:cratering-efficiency}. Although the negative correlation between $\pi_{\rm v}$ and $\pi_{\rm 2}$ can clearly be confirmed, the data in Fig.~\ref{fig:cratering-efficiency}(a) show considerable scattering. The reason for this data scattering is rather obvious. We varied both $\theta$ and $\varphi$ in this experiment whereas these factors are not taken into account in the dimensionless numbers $\pi_{\rm v}$ and $\pi_{\rm 2}$. To improve the quality of data collapse in the scaling plot, we have to modify the dimensionless numbers by considering $\theta$ and $\varphi$. First, we consider the effect of $\varphi$. According to \citet{Gault:1978}, the crater volume $V_{\rm c}$ decreases as the impact angle $\varphi$ becomes small, obeying the factor $\sin\varphi$. In general, $\sin\varphi$ dependence comes from the contribution of the normal component of impact inertia to the crater formation process. Therefore, we modify $\pi_{\rm 2}$ by using the factor $\sin\varphi$ as,
\begin{equation}
 \pi'_{\rm 2}=\frac{gD_{\rm i}}{v_{\rm i}^2 \sin\varphi}.
 \label{eq:pi2'}
\end{equation}

Figure~\ref{fig:cratering-efficiency}(b) shows the relation between $\pi_{\rm v}$ and $\pi'_{\rm 2}$. Although the data of identical $\theta$ seem to collapse onto scaling relations (straight lines in log-log plot), $\theta$ dependence of $\pi_{\rm v}$ remains as data scattering. However, one can confirm that the scaling exponent ($\simeq 0.52$; slope in Fig.~\ref{fig:cratering-efficiency}(b)) is almost independent of $\theta$. Namely, the scaling law can be written as,
\begin{equation}
 \pi_{\rm v}=f(\theta){\pi'_{\rm 2}}^{-a},
 \label{eq:piv}
\end{equation}
 where $f$ is a certain dimensionless function and $a=0.52$ is a scaling exponent. To obtain a specific functional form of $f$, $f(\theta)$ at $\pi'_{\rm 2}=1$ is scaled by $\cos\theta$ as shown in the inset of Fig.~\ref{fig:cratering-efficiency}(c). By assuming the power-law form, scaling parameters can be obtained from the least square fitting to the data as shown in the inset of Fig.~\ref{fig:cratering-efficiency}(c). Here, we empirically employ a variable $\cos\theta$ to simply recover the horizontal case ($\theta=0$) by $\cos\theta=1$. In other words, we use $\cos\theta$ because it becomes unity in the standard case $\theta=0^{\circ}$. The obtained empirical scaling relation is expressed as,
\begin{equation}
 f(\theta)=0.14(\cos\theta)^{4.5}.
 \label{eq:f}
\end{equation}
Based on this relation, $\pi_{\rm v}$ is modified using the factor $\cos\theta$ as,
\begin{equation}
 \pi'_{\rm v}=\frac{\rho_{\rm t}V_{\rm c}}{m_{\rm i}{(\cos\theta)}^{4.5}}.
 \label{eq:piv'_difinition}
\end{equation}

 Figure~\ref{fig:cratering-efficiency}(c) shows the relation between $\pi'_{\rm v}$ and $\pi'_{2}$. Finally, all data are collapsed onto a unified scaling relation. The scaling law including the effects of $\sin\varphi$ and $\cos\theta$ is finally written as
\begin{equation}
 \frac{\rho_{\rm t}V_{\rm c}}{m_{\rm i}}=0.14(\cos\theta)^{4.5}\left[\frac{gD_{\rm i}}{v_{\rm i}^2 \sin\varphi}\right]^{-0.52}.
 \label{eq:piv'}
\end{equation}
 The relation between $\pi'_{2}$ and $\pi'_{\rm v}$ obeys power law form with nontrivial scaling exponent $0.52$ (obtained in the fitting in Fig.~\ref{fig:cratering-efficiency}(c)) and scaling coefficient $0.14(\cos\theta)^{4.5}$. Eq.~(\ref{eq:piv'}) can be rewritten as 
\begin{equation}
\pi_V=0.14(\cos\theta)^{4.5}(\sin\theta)^{0.52}\pi_2^{-0.52}.
\label{eq:piv-simple}
\end{equation}
In this analysis, the common exponent value $0.52$ is used for both $\pi_{\rm 2}$ and $\sin\varphi$ to reduce the number of free fitting parameters in the scaling model. Nevertheless, the quality of data collapse by the scaling is excellent. Alternatively, the normal component of impact velocity ($v_i \sin\varphi$) could be an important parameter~\citep{Chapman:1986}. To check this form, we also tried the data collapse by using $\pi'_{\rm 2}=gD_{\rm i}/(v_{\rm i}\sin\varphi)^2$. However, the quality of data collapse was better when Eq.~(\ref{eq:pi2'}) was used for the definition of $\pi'_{\rm 2}$ (see Fig.~\ref{fig:cratering-efficiency}(b)). Namely, the scaling of Eq.~(\ref{eq:piv'}) is the best one to reasonably fit the experimental data. 

To discuss the physical meaning of the obtained scaling, a coupling parameter called the point-source measure is defined as, 
\begin{equation}
C_{\rm p}=D_{\rm i} v_{\rm i}^{\mu}\rho_{\rm i}^{1/3}, 
\label{eq:Cpsm}
\end{equation}
where $\rho_{\rm i}$ is the density of projectile. The physical meaning of $C_{\rm p}$ can be evaluated by the value of exponent $\mu$. In the limit of $\mu=1/3$ (or $2/3$), $C_{\rm p}^3$ corresponds to the impact momentum (or kinetic energy). Usually, the value of $\mu$ distributes between $1/3$ and $2/3$ depending on the experimental conditions. In addition, the value of $\mu$ can be related to the scaling exponent on $\pi_2$. Specifically, 
the scaling exponent is expressed as $0.52=3\mu/(2+\mu)$~(e.g. \citep{Holsapple:1993}). From this relation, $\mu=0.42$ is obtained. In general, the value of $\mu$ strongly depends on target properties such as porosity and internal friction \citep{Wunnemann:2006,Elbeshausen:2009}. For example, the representative values $\mu=0.41$ and $\mu=0.55$ are obtained for dry sand target and nonporous target, respectively~\citep{Holsapple:1987,Gault:1982}. The $\mu$ value obtained in this experiment agrees with the previous study using dry sand target. 
This value ($\mu=0.42$) indicates that the current impact situation is closer to the momentum-scaling limit ($\mu=1/3$) than energy-scaling limit ($\mu=2/3$). In other words, the momentum transfer could be more important than the energy transfer for the dissipative impact using sand target like this experiment. 

\begin{figure}
\centering
\includegraphics[clip,width=1.0\linewidth]{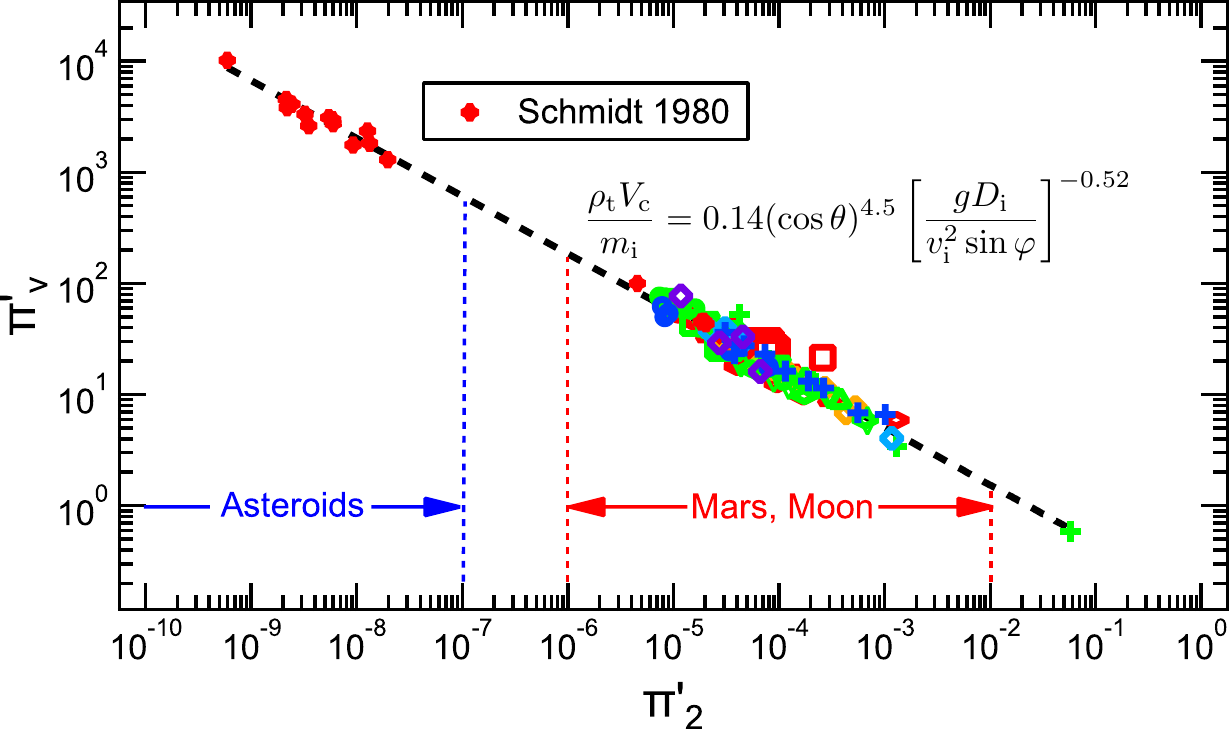}
\caption{
Universal scaling relation valid for both this study and previous experiment~\citep{Schmidt:1980}.
The $\pi'_{\rm 2}$ data come from \citet{Schmidt:1980} and this study. All data obey the scaling (Eq.~(\ref{eq:piv'})). The additional data point with the largest $\pi_{\rm 2}$ value (not shown in the previous plots) originates from the free-fall (very low-speed, $v_{\rm i} \simeq 1$~m~s$^{-1}$) collision. 
The red range represents the range of an impact with $D_{\rm i}=0.1$--$10$~km colliding with a Moon-or-Mars scale target (${\it g}\sim10^0$ m~s$^{-2}$) at $v_{\rm i}=1$--$10$~km~s$^{-1}$. On the other hand, the blue range represents the impact with $D_{\rm i}=0.1$--$10$~m colliding with a small asteroid ($g\sim10^{-4}$~m~s$^{-2}$) at $v_{\rm i}=0.1$--$10$~km~s$^{-1}$. Color and symbol codes used in this plot are identical to those in Fig.~\ref{fig:cratering-efficiency}.
}
\label{fig:consistency-with-the-previous-study-(piv)}
\end{figure}

Actually, the form of Eq.~(\ref{eq:piv'}) is similar to the scaling obtained by the high-speed impact ($v_{\rm i}\sim 10^3$~m~s$^{-1}$) into a sand target~\citep{Schmidt:1980}. \citet{Schmidt:1980} performed the experiment of normal impact onto a flat surface under the vacuum condition. This situation simply corresponds to $\sin\varphi=\cos\theta=1$ in Eq.~(\ref{eq:piv'}). Figure~\ref{fig:consistency-with-the-previous-study-(piv)} shows the consistency between the current result and the previous study~\citep{Schmidt:1980}.
In Fig.~\ref{fig:consistency-with-the-previous-study-(piv)}, the data reported in \citet{Schmidt:1980} are plotted as well as the current experimental results. Specifically, the data taken with Ottawa Flintshot sand (17 shots) in \citet{Schmidt:1980} are plotted. While the experimental details such as used materials and measuring methods are different between our experiment and \citet{Schmidt:1980}, the mutual consistency of these data can be checked by simply plot both data in the same graph. In addition, one data point of a very low-speed ($v_{\rm i} \simeq 1$~m~s$^{-1}$) impact (the largest $\pi'_2$ data) is added in this plot (not shown in the previous plots). Although this data point comes from the free-fall impact to the flat surface, it completely obeys the scaling. The excellent data collapse shown in Fig.~\ref{fig:consistency-with-the-previous-study-(piv)} suggests that Eq.~(\ref{eq:piv'}) is a universal scaling relation independent of the experimental conditions such as impact-velocity range and ambient pressure. The weak point of this study is the too low impact speed to mimic large-scale astronomical impacts. According to \citet{Yamashita:2009}, typical sound speed in Toyoura sand is $\simeq 250$~m~s$^{-1}$. This value is larger than the impact speed in the current experiment. Thus, underlying physics of the current experiment could be different from that of the hyper-velocity impact. However, the current result is fully consistent with the hyper-velocity impact experiment. \citet{Schmidt:1980} performed the hyper-velocity-impact experiment under the vacuum condition while our experiment is conducted with low impact speed under the atmospheric pressure conditions. Nevertheless, both results obey the identical scaling law. Since the scaling law is written in dimensionless form and valid for both hyper-velocity and low-velocity impacts, we expect that the scaling is valid also for the large-scale phenomena. The scaling law obtained by \citet{Schmidt:1980} is very robust. In this study, we expand the scaling form by using two factors: $\cos\theta$ and $\sin\varphi$.

\begin{figure}
\centering
\includegraphics[clip,width=1.0\linewidth]{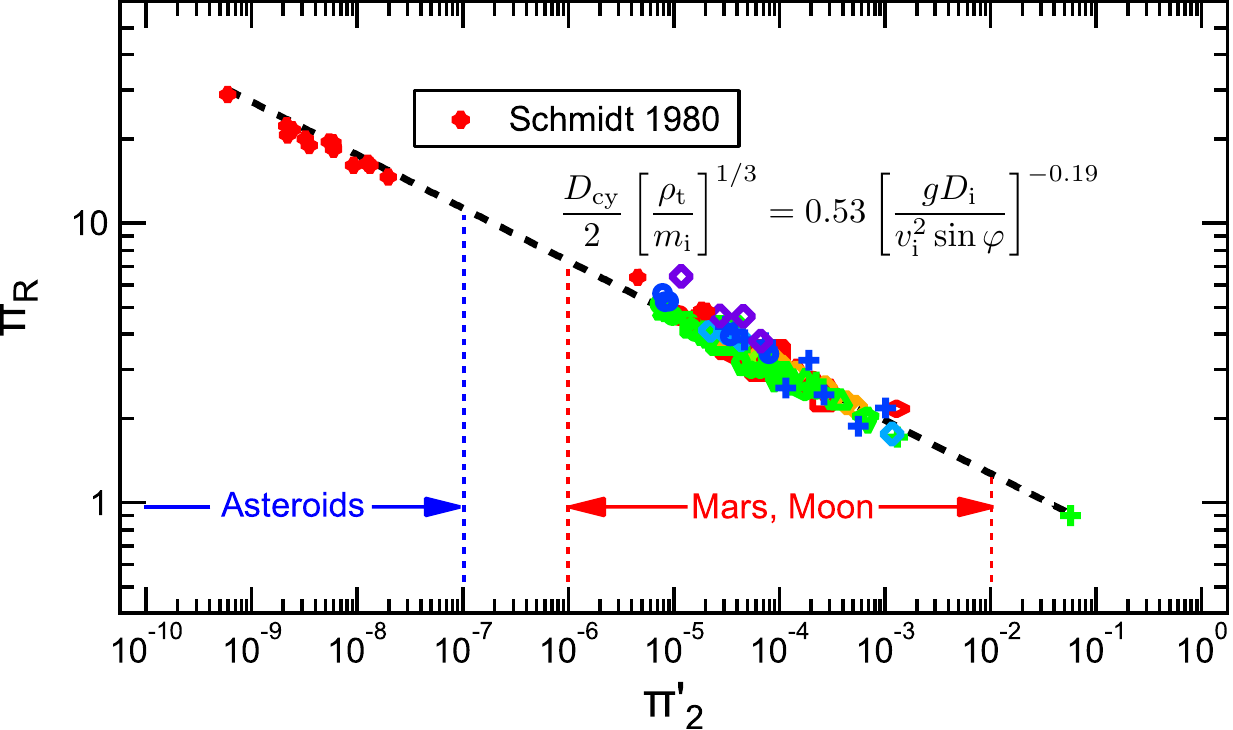}
\caption
{
Crater diameter scaling using dimensionless numbers $\pi_{\rm R}$ and $\pi'_{\rm 2}$. The indicated regions and codes for color and symbols are identical to those shown in Fig.~\ref{fig:consistency-with-the-previous-study-(piv)}.
}
\label{fig:consistency-with-the-previous-study-(piR)}
\end{figure}

By the same protocol, crater diameter $D_{\rm cy}$ can also be scaled. Since $D_{\rm cy}$ is independent of $\theta$, we can directly scale the non-dimensionalized crater radius,
\begin{equation}
\pi_{\rm R}=\frac{D_{\rm cy}}{2}\left[\frac{\rho_{\rm t}}{m_{\rm i}}\right]^{\frac{1}{3}},
 \label{eq:piR_definition2}
\end{equation}
using $\pi'_{\rm 2}$ as shown in Fig.~\ref{fig:consistency-with-the-previous-study-(piR)}. We can clearly confirm the scaling relation, 
\begin{equation}
 \frac{D_{\rm cy}}{2}\left[\frac{\rho_{\rm t}}{m_{\rm i}}\right]^{\frac{1}{3}}=0.53\left[\frac{gD_{\rm i}}{v_{\rm i}^2 \sin\varphi}\right]^{-0.19},
 \label{eq:piR}
\end{equation}
 where the coefficient $0.53$ and exponent $0.19$ are computed by the least square fitting to all data. In Fig.~\ref{fig:consistency-with-the-previous-study-(piR)}, data from this study and from \citet{Schmidt:1980} are plotted just like Fig.~\ref{fig:consistency-with-the-previous-study-(piv)}. Again, the excellent agreement between this study and \citet{Schmidt:1980} can be confirmed. 

 The ranges of $\pi'_{\rm 2}$ shown in Figs.~\ref{fig:consistency-with-the-previous-study-(piv)} and \ref{fig:consistency-with-the-previous-study-(piR)} correspond to those for typical astronomical impacts on planets (or satellites like Mars and Moon), or asteroids (like an asteroid 162173 Ryugu). The value of $\pi'_{\rm 2}$ ranges in $10^{-6}$--$10^{-2}$ when we consider the gravity corresponding to Moon or Mars ($g\sim 10^0$~m~s$^{-2}$), $D_{\rm i}=0.1$--$10$~km, and $v_{\rm i}=1$--$10$~km~s$^{-1}$.  On the other hand, $\pi'_{\rm 2}=10^{-13}$--$10^{-7}$ is obtained when we consider small-asteroids-level gravity $g=10^{-4}$~m~s$^{-2}$, $D_{\rm i}=0.1$--$10$~m, and $v_{\rm i}=0.1$--$10$~km~s$^{-1}$. As shown in Figs.~\ref{fig:consistency-with-the-previous-study-(piv)} and \ref{fig:consistency-with-the-previous-study-(piR)}, the current experimental result corresponds to the relatively larger-scale impacts while the previous study~\citep{Schmidt:1980} actually simulates the asteroid-scale impacts. Namely, in terms of dimensional analysis, the obtained scaling law covers from asteroid scale to planet scale. The scaling law is satisfied over eight orders of magnitude in $\pi'_{\rm 2}$ as shown in Figs.~\ref{fig:consistency-with-the-previous-study-(piv)} and \ref{fig:consistency-with-the-previous-study-(piR)}. Although we have assumed the scale-free nature of the cratering phenomena based on the similarity law, the absolute impact speed used in this experiment is small. However, the excellent data collapse shown in Figs.~\ref{fig:consistency-with-the-previous-study-(piv)} and \ref{fig:consistency-with-the-previous-study-(piR)} strongly suggests the wide applicability of the scaling laws.

\subsection{Scaling of the crater aspect ratios}
\begin{figure}
\centering
\includegraphics[clip,width=1.0\linewidth]{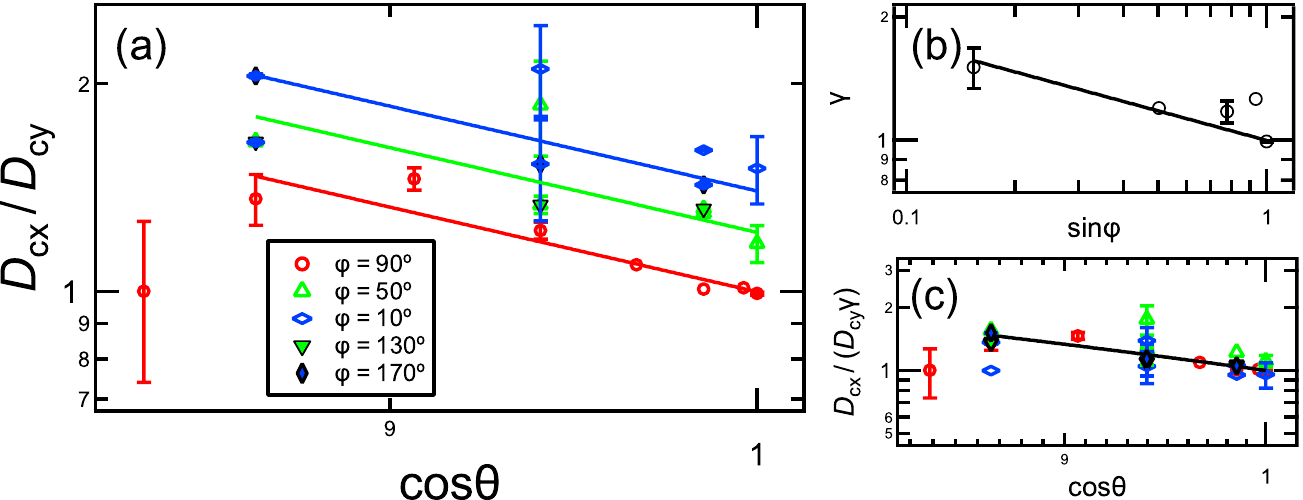}
\caption{
 Scaling of the aspect ratio of the crater $D_{\rm cx}$/$D_{\rm cy}$. (a)~The relation between $D_{\rm cx}$/$D_{\rm cy}$ and $\cos\theta$. Note that the identical color indicates the data of the same incident angle relative to $\varphi=90^{\circ}$. (b)~$\gamma$ vs $\sin\varphi$, and (c)~$D_{\rm cx}$/($D_{\rm cy}\gamma)$ vs $\cos\theta$ are plotted to obtain the scaling form of Eq.~(\ref{eq:aspect-ratio}). Note that all the plots are shown in log-log style. The range of impact speed is $10<v_{\rm i}<100$~m~s$^{-1}$.
}
\label{fig:aspect-ratio}
\end{figure}

Thus far, we have focussed on the scaling analyses of the crater volume and diameter. The obtained scaling laws are consistent wth previous study. However, to understand the effect of $\theta$ and $\varphi$ on the crater asymmetry, we should analyze the aspect ratio of craters as well. Indeed, the crater aspect ratio can easily be measured even in natural craters, and its characterization is crucial to extract the useful information about the impact conditions for natural craters. Therefore, two kinds of aspect ratios $D_{\rm cx}/D_{\rm cy}$ and $H_{\rm c}/D_{\rm cy}$ are analyzed in this subsection. Here we assume that these aspect ratios are independent of the scale of impact, i.e., the crater morphology should be similar independent of $E$. That is, we assume that the energy scaling shown in Fig.~\ref{fig:Energy-scalings} is universal in wider range. This assumption is supported by the widely holding scaling relations shown in Figs.~\ref{fig:consistency-with-the-previous-study-(piv)} and \ref{fig:consistency-with-the-previous-study-(piR)}. Then, the aspect ratios should depend only on $\theta$ and $\varphi$. In Fig.~\ref{fig:aspect-ratio}(a), the relation between $D_{\rm cx}$/$D_{\rm cy}$ and $\cos\theta$ is shown. As seen in Fig.~\ref{fig:aspect-ratio}(a), $D_{\rm cx}$/$D_{\rm cy}$ decreases with $\cos\theta$ for each $\varphi$. This trend is qualitatively consistent with previous studies~\citep{Hayashi:2017,Aschauer:2017}. In addition, we can confirm that $D_{\rm cx}/D_{\rm cy}$ satisfies the relation $[D_{\rm cx}/D_{\rm cy}](\varphi)\simeq [D_{\rm cx}/D_{\rm cy}](180^{\circ} -\varphi)$. In other words, the data of $\varphi=50^{\circ}$ (or $\varphi=10^{\circ}$) and $\varphi=130^{\circ}=180^{\circ}-50^{\circ}$ (or $\varphi=170^{\circ}$) obey the identical scaling as shown in Fig.~\ref{fig:aspect-ratio}(a). Namely, $\sin\varphi$ is a relevant parameter to analyze the data. From Fig.~\ref{fig:aspect-ratio}(a), an empirical scaling form $D_{\rm cx}$/$D_{\rm cy}=\gamma(\sin\varphi)[\cos\theta]^{-3}$ is obtained. Here, $\gamma$ is a certain dimensionless function of $\sin\varphi$. To determine the form of $\gamma(\sin\varphi)$, the relation between $\gamma$ and $\sin\varphi$ is shown in Fig.~\ref{fig:aspect-ratio}(b). As a result, we empirically obtain $\gamma=[\sin\varphi]^{-0.2}$. Thus, we finally obtain the relation,
\begin{equation}
 \frac{D_{\rm cx}}{D_{\rm cy}}=\left[\sin\varphi\right]^{-0.2}\left[\cos\theta\right]^{-3}.
 \label{eq:aspect-ratio}
\end{equation}
The corresponding normalized scaling relation is plotted in Fig.~\ref{fig:aspect-ratio}(c). 

\begin{figure}
\centering
\includegraphics[clip,width=1.0\linewidth]{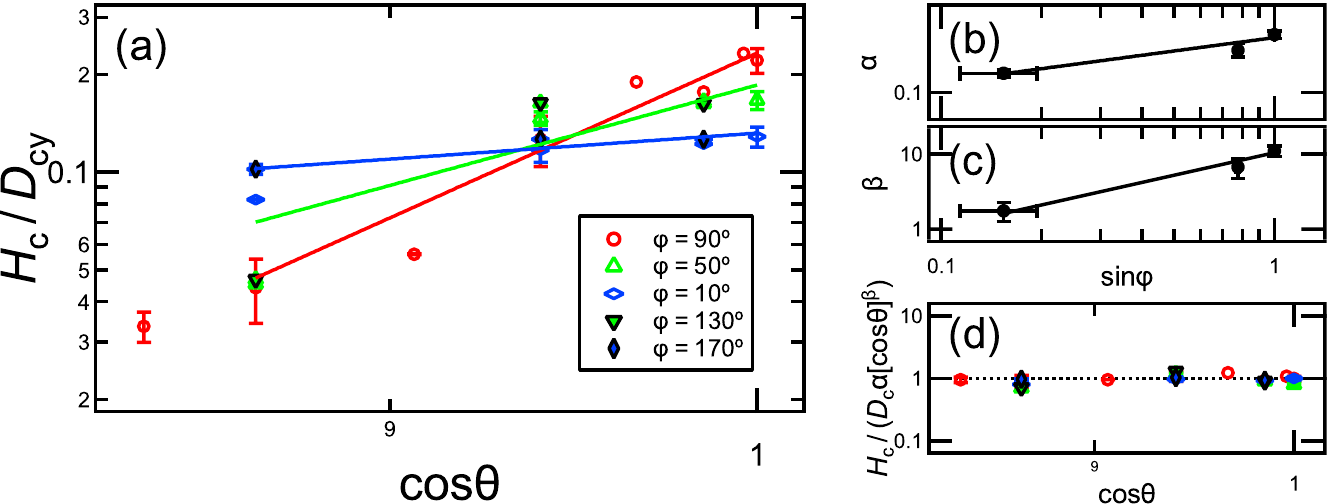}
\caption{
Scaling of the depth-diameter ratio $H_{\rm cx}$/$D_{\rm cy}$. (a)~The relation between $H_{\rm c}/D_{\rm cy}$ and $\cos\theta$. (b) Impact angle ($\sin\varphi$) dependence of the scaling coefficient $\alpha$. (c) Impact angle ($\sin\varphi$) dependence of the scaling exponent $\beta$. (d) Final scaling plot on the basis of Eq.~(\ref{eq:depth-diameter-ratio}). Note that all the plots are shown in log-log style.
}
\label{fig:depth-diameter}
\end{figure}

Similarly, the depth-diameter ratio $H_{\rm c}/D_{\rm cy}$ is scaled by $\cos\theta$ and $\sin\varphi$. The relation between $H_{\rm c}/D_{\rm cy}$ and $\cos\theta$ is displayed in Fig.~\ref{fig:depth-diameter}(a). Although $H_{\rm c}/D_{\rm cy}$ seems to be scaled by $\cos\theta$, both the scaling coefficient and exponent depend on $\sin\varphi$. The corresponding scaling form is written as $H_{\rm c}/D_{\rm cy}=\alpha(\sin\varphi)[\cos\theta]^{\beta(\sin\varphi)}$ ($\alpha$ and $\beta$ are dimensionless functions of $\sin\varphi$). Therefore, the relations between $\alpha$ vs $\sin\varphi$ and $\beta$ vs $\sin\varphi$ are plotted in Figs.~\ref{fig:depth-diameter}(b) and (c), respectively. As a consequence, we empirically obtain two relations: $\alpha=0.2[\sin\varphi]^{0.3}$ and $\beta=10\sin\varphi$. Thus, the scaling for $H_{\rm c}/D_{\rm cy}$ is written as,
\begin{equation}
 \frac{H_{\rm c}}{D_{\rm cy}}=0.2\left[\sin\varphi\right]^{0.3}\left[\cos\theta\right]^{10\sin\varphi}.
 \label{eq:depth-diameter-ratio}
\end{equation}
The corresponding scaling plot is shown in Fig.~\ref{fig:depth-diameter}(d).

 These scaling laws are obtained empirically. Moreover, the effects of $\theta$ and $\varphi$ are not independent in the scaling of $H_c/D_{cy}$ (Eq.~(\ref{eq:depth-diameter-ratio})). Whereas these relations are dimensionless, physical basis on these scaling laws is not very firm. Furthermore, the scale-free (similarity) assumption for the aspect ratios might not be held in large scale. The relative population of the elliptic craters actually depends on the scale of craters~\citep{Collins:2011}. Therefore, the validity of these relations has to be checked by observational data. This is a crucial future work.

\subsection{Possible application to astronomical impact cratering}
Thus far, we have derived scaling laws for the crater dimensions and aspect ratios (Eqs.~(\ref{eq:piv'}), (\ref{eq:piR}), (\ref{eq:aspect-ratio}), and (\ref{eq:depth-diameter-ratio})). Since these relations are dimensionless, they are scale free and basically applicable to large-scale phenomena (astronomical impact cratering). Therefore, in this section, we consider a possible way to  estimate the impact conditions such as impact angle, slope of target terrain, impact velocity, etc. from the observable crater dimensions. The parameters included in the scaling laws are $D_{\rm i}$, $\rho_{\rm i}$, $v_{\rm i}$, $\rho_{\rm t}$, $\rm g$, $\theta$, and $\varphi$. Among them, $D_{\rm i}$, $\rho_{\rm i}$, and $v_{\rm i}$ can be combined to a coupling parameter $C_{\rm p}$ (Eq.~(\ref{eq:Cpsm})). In addition, $\rho_{\rm t}$ and $g$ can be estimated from the observation of the target body. Therefore, we have three unknown parameters ($C_{\rm p}$, $\theta$, and $\varphi$) and four scaling laws~(Eqs.~(\ref{eq:piv'}), (\ref{eq:piR}), (\ref{eq:aspect-ratio}), and (\ref{eq:depth-diameter-ratio})). Namely, from the four observables, $V_{\rm c}$, $D_{\rm cy}$, $D_{\rm cx}$, and $H_{\rm c}$, one can obtain the three value, $C_{\rm p}$, $\theta$, and $\varphi$. For fresh craters, $\theta$ can also be estimated from the observation. Then, four parameters,  $D_{\rm i}$, $\rho_{\rm i}$, $v_{\rm i}$, and $\varphi$ (three ingredients of $C_{\rm p}$ and the impact angle) can be computed. However, we have to be careful when the scaling laws are used to analyze the astronomical impact craters. For instance, crater relaxation caused by impact-induced seismic shaking should affect the shape of old craters on small bodies~\citep{Richardson:2004,Richardson:2005,Katsuragi:2016,Tsuji:2018,Tsuji:2019}. For small bodies like asteroids, impact-induced resurfacing~\citep{Yamada:2016} could also affect the surface terrain. Besides, small impacts impinging the wall of a large crater also relaxes the crater shape on relatively large target bodies~\citep{Soderblom:1970}. Actually, the relaxation of lunar craters have been analyzed by this type of gradual erosion~\citep{Fassett:2014}. On the surface of large astronomical bodies, tectonic deformation and ejecta deposition from the adjacent large-scale impacts could modify the crater shape significantly. Thus, it is difficult to separate such additional modification effects from the instantaneous $\theta$ and $\varphi$ effects, particularly for old craters. Therefore, the scaling laws can only be applied to fresh craters. In addition, the aspect ratio $D_{\rm cx}/D_{\rm cy}$ could depend on the crater scale. Particularly, this effect becomes significant in very large craters~\citep{Collins:2011}. Besides, we neglect the effect of melting and shock wave propagation induced by the hypervelocity impact.

\subsection{Future problems}
In order to verify the validity of the obtained scaling laws, statistical analysis of a large number of astronomical craters could be useful. Specifically, the effect of the inclination angle $\theta$ on the crater shapes ($D_{\rm c}/D_{\rm cy}$ and $H_{\rm c}/D_{\rm cy}$) could be statistically verified by examining the relation between fresh crater shapes and the local inclination angles on e.g., Moon or Mars on which abundant crater records can be found. The relations of Eqs.~(\ref{eq:aspect-ratio}) and (\ref{eq:depth-diameter-ratio}) are obtained on the basis of laboratory experiment. The statistical analysis could be helpful to improve these relations by combining experimental and observational results. 

In this experiment, impact speed is slower than the target sound speed. Although the scaling laws derived in this study is completely consistent with the previously performed hyper-velocity-impact experiment, the hyper-velocity impacts with various $\theta$ and $\varphi$ should be carried out to directly obtain the conclusive scaling laws. Moreover, we only use Toyoura sand for the target material. Thus, we cannot evaluate the effect of target mechanical properties such as internal friction and grain size. Obviously, actual planetary/asteroidal surfaces have large variations in terms of mechanical properties. Therefore, systematic experiments  with various target materials have to be conducted to properly understand the actual surface processes on astronomical bodies.

The current oblique-impact experiment neglects the effect of inclination around $X$ axis. In this study, all the angle variations are defined around $Y$ axis. This is the reason why $D_{\rm cy}$ is independent of $\theta$. To consider the truly 3D oblique impact, the incident angle around $X$ axis should also be taken into account. Furthermore, spin of projectile could also affect the cratering dynamics. However, these effects are probably limited since the former could only affect the principal axis directions of the elongated crater shape and the latter could only increase the effective impact energy. Thus, we focus on the simpler case in this study.

In addition, to reveal the crater formation process in more detail, it is necessary to develop the scaling laws for the ejecta velocity and the timescale of crater formation. The detail analysis of rebound velocity and the timescale of rebound are also interesting future problems. By the preliminary analysis, we find that the rebound timescale significantly depends on $\varphi$. The rebound dynamics depending on $\varphi$ could be a key factor to understand the physics of general impact cratering phenomena.

\section{Conclusion}
In this study, oblique-impact experiments onto an inclined granular layer were performed for understanding the crater formation process and obtaining the scaling laws including the effects of both impact angle $\varphi$ and inclination angle of the target surface $\theta$. From the classification diagram based on the type of crater-wall collapse, we found that the scale of collapse of upper wall on transient crater cavity depends mainly on $\theta$. As a consequence of the collapse, the final crater dimensions differently depend on the impact kinetic energy $E$. While the crater length $D_{\rm cx}$ is an increasing function of $E$, crater width $D_{\rm cy}$ is independent of $E$. And, the crater depth $H_{\rm c}$ and volume $V_{\rm c}$ decrease as $E$ increases. To obtain the universal scaling for the crater dimensions, parameters were non-dimensionalized and analyzed on the basis of $\Pi$-group scaling. In addition to the conventional dimensionless numbers ($\pi_{\rm v}=\rho_{\rm t}V_{\rm c}/m_i$, $\pi_{\rm R}=(D_{\rm cy}/2)(\rho_{\rm t}/m_{\rm i})^{1/3}$, and $\pi_{\rm 2}=gD_i/v_i^2$), we took into account the effects of two angle factors: $\sin\varphi$ and $\cos\theta$. As a result, we found that the normalized crater volume $\pi'_{\rm v}=\rho_{\rm t}V_{\rm c}/m_i[\cos\theta]^{4.5}$ and $\pi_{\rm R}$ are scaled by the modified gravity parameter $\pi'_{\rm 2}=gD_i/(v_i^2\sin\varphi)$ as written in Eqs.~(\ref{eq:piv'}) and (\ref{eq:piR}). Because the obtained scaling laws are fully consistent with the previous study of hyper-velocity impacts, we consider the scaling laws are universal and applicable to the astronomical-scale phenomena. Besides, the crater aspect ratios $D_{\rm cx}/D_{\rm cy}$ and $H_{\rm c}/D_{\rm cy}$ were scaled by $\sin\varphi$ and $\cos\theta$ (Eqs.~(\ref{eq:aspect-ratio}) and (\ref{eq:depth-diameter-ratio})). Now, we have four scaling laws to estimate the impact conditions from the crater morphology. By assuming some quantities, impact conditions (e.g., $D_{\rm i}$, $\rho_{\rm i}$, $v_{\rm i}$, $\varphi$ etc.) can be estimated from the crater dimensions by using these scaling laws. Namely, we have successfully obtained a set of scaling laws that are useful to analyze the natural impact craters. However, the validity of the obtained scaling should be checked with various (projectile/target) materials and impact speeds to strengthen their applicability to various impact situations. 

\section*{Acknowledgement}
The authors acknowledge R. Yamaguchi and H. Niiya for the development of the impact apparatus and helpful discussion. This work has been supported by JSPS KAKENHI Grant No.~18H03679. 

\section*{Data availability}
All the experimental raw data are tabulated in the supplementary file (all-data.csv). The file includes the data set of impact conditions $\theta$, $\varphi$, $v_{\rm i}$ and $m_{\rm i}$, and the measured crater dimensions: $D_{\rm cx}$, $D_{\rm cy}$, $H_{\rm c}$, and $V_{\rm c}$.

\bibliographystyle{elsarticle-harv} 

\bibliography{oblique_impact}

\end{document}